%% file: main.tex
\documentclass[preprint,12pt]{elsarticle}

\usepackage{xcolor}

\usepackage{graphicx}% Include figure files
\usepackage{xcolor}
\usepackage{longtable}
\usepackage{caption}
\usepackage{float}
\usepackage[caption=false]{subfig}
\usepackage[utf8]{inputenc}
\usepackage[T1]{fontenc}
\usepackage{hyperref}
\usepackage{multirow}
\usepackage{lscape}
\usepackage{makecell}

\usepackage{tikz}
\usetikzlibrary{3d}
\usepackage{pgfplots}
\tikzset{my_rectangle/.style={draw,rectangle, minimum width=4cm, minimum height=4cm}}
\usepackage{amsmath} %permette di scrivere formule matemtiche
\usepackage{amssymb} %permette di inserire simboli
\usepackage{amsthm} %permette di inserire teoremi pre-struttrati
\usepackage{amscd} %diagrammi commutativi o di Venn
\usepackage{braket} %per braket \Braket{"testo del bra"|"testo del ket"}
\usepackage{empheq} %evidenziare formule
\usepackage{bm} %lettere greche in grassetto \bm{"lettera"}
\usepackage{mathabx}
\usepackage{physics}
\usepackage[normalem]{ulem}
\biboptions{numbers,sort&compress}
\pgfplotsset{compat=1.18} 

\journal{.....}

\begin{document}

\begin{frontmatter}

%% Title, authors and addresses

%% use the tnoteref command within \title for footnotes;
%% use the tnotetext command for theassociated footnote;
%% use the fnref command within \author or \address for footnotes;
%% use the fntext command for theassociated footnote;
%% use the corref command within \author for corresponding author footnotes;
%% use the cortext command for theassociated footnote;
%% use the ead command for the email address,
%% and the form \ead[url] for the home page:
%% \title{Title\tnoteref{label1}}
%% \tnotetext[label1]{}
%% \author{Name\corref{cor1}\fnref{label2}}
%% \ead{email address}
%% \ead[url]{home page}
%% \fntext[label2]{}
%% \cortext[cor1]{}
%% \affiliation{organization={},
%%             addressline={},
%%             city={},
%%             postcode={},
%%             state={},
%%             country={}}
%% \fntext[label3]{}

%\title{Analytically-enabled super resolution and local (de)refinement can significantly benefit discontinuous coefficient PDEs: the Poisson-Boltzmann Equation case NextGenPB}

\title{NextGenPB: an analytically-enabled super resolution and 
local (de)refinement Poisson-Boltzmann Equation solver}

%\title{Analytically-enabled super resolution , combining analytical and numerical solutions to keep the and and approach}

%% use optional labels to link authors explicitly to addresses:
%% \author[label1,label2]{}
%% \affiliation[label1]{organization={},
%%             addressline={},
%%             city={},
%%             postcode={},
%%             state={},
%%             country={}}
%%
%% \affiliation[label2]{organization={},
%%             addressline={},
%%             city={},
%%             postcode={},
%%             state={},
%%             country={}}

\author[inst1,inst2,inst3]{Vincenzo Di Florio}
\author[inst4]{Patrizio Ansalone}
\author[inst1]{Sergii V. Siryk}
\author[inst10]{Sergio Decherchi}
\author[inst3]{Carlo de Falco}
\author[inst1]{Walter Rocchia}

\affiliation[inst1]{organization={CONCEPT Lab, Fondazione Istituto Italiano di Tecnologia},%Department and Organization
            addressline={Via E. Melen 83}, 
            city={Genova},
            postcode={16152}, 
            % state={State One},
            country={Italy}}

\affiliation[inst2]{organization={Department of Mathematical Sciences, Politecnico di Torino},%Department and Organization
            addressline={ Corso Duca degli Abruzzi, 24}, 
            city={Torino},
            postcode={10129}, 
            % state={},
            country={Italy}}

\affiliation[inst3]{organization={MOX Laboratory, Department of Mathematics, Politecnico di Milano},%Department and Organization
            addressline={Piazza Leonardo Da Vinci, 32}, 
            city={Milano},
            postcode={20133}, 
            % state={},
            country={Italy}}

\affiliation[inst4]{organization={ Advanced Materials Metrology and Life Sciences, Istituto Nazionale di Ricerca Metrologica},%Department and Organization
            addressline={Strada delle Cacce, 91}, 
            city={Torino},
            postcode={10135}, 
            %state={},
            country={Italy}}

\affiliation[inst10]{organization={Data Science and Computation, Fondazione Istituto Italiano di Tecnologia},%Department and Organization
            addressline={Via Morego 30}, 
            city={Genova},
            postcode={16163}, 
            % state={State One},
            country={Italy}}

\begin{abstract}
%A prevalent viewpoint regarding the numerical solution of Partial Differential Equations (PDEs) suggests that enhancing accuracy solely relies on increasing discretization resolution. While this method is effective and broadly applicable, it comes with associated complexities, computational costs, and memory demands. This work proposes an alternative approach integrating analytical derivations with conventional finite element methods to maximize accuracy at a given resolution while minimizing the computational cost-to-accuracy ratio.

%A trivial approach for increasing the accuracy of the numerical solution of Partial Differential Equations (PDEs) consists in increasing discretization resolution. While effective and broadly applicable, this approach has associated complexities, computational costs, and memory demands. This work proposes an alternative approach that integrates analytical derivations with conventional finite element methods to maximize accuracy at a given resolution while minimizing the computational cost-to-accuracy ratio.

The Poisson-Boltzmann equation (PBE) is a relevant partial differential equation commonly used in biophysical applications to estimate the electrostatic energy of biomolecular systems immersed in electrolytic solutions. A conventional mean to improve the accuracy of its solution, when grid-based numerical techniques are used, consists in increasing the resolution,  locally or globally. This, however, usually entails higher complexity, memory demand and computational cost. Here, we introduce NextGenPB, a linear PBE, adaptive-grid, FEM solver that leverages analytical calculations to maximize the accuracy-to-computational-cost ratio. 
Indeed, in NextGenPB (aka NGPB), analytical corrections at the surface of the solute enhance the solution's accuracy without requiring grid adaptation. This leads to more precise estimates of the electrostatic potential, fields, and energy at no perceptible additional cost.
Also, we apply computationally efficient yet accurate boundary conditions by taking advantage of local grid de-refinement. 
To assess the accuracy of our methods directly, we expand the traditionally available analytical case set to many non-overlapping dielectric spheres. Then, we use an existing benchmark set of real biomolecular systems to evaluate the energy convergence concerning grid resolution.  
Thanks to these advances, we have improved state-of-the-art results and shown that the approach is accurate and largely scalable for modern high-performance computing architectures. Lastly, we suggest that the presented core ideas could be instrumental in improving the solution of other partial differential equations with discontinuous coefficients.

\end{abstract}

%%Graphical abstract
% \begin{graphicalabstract}
% \includegraphics{grabs}
% \end{graphicalabstract}

%%Research highlights

% \begin{highlights}

% \item An analytically-enhanced approach for solving the Poisson-Boltzmann equation

% \item Introduction of an analytical benchmark constituted by many non-overlapping spheres, enabling rigorous validation of PBE solvers

% \item Exploitation of accurate molecular surface descritpion provided by the NanoShaper software tool

% \item large systems

% \item remarkable trade off between accuracy and computational cost

% \end{highlights}

\begin{keyword}
%% keywords here, in the form: keyword \sep keyword
Poisson-Boltzmann \sep Continuum electrostatics \sep Finite-Element method \sep Analytical PDE solutions \sep PBE solver
%% PACS codes here, in the form: \PACS code \sep code
\PACS 0000 \sep 1111
%% MSC codes here, in the form: \MSC code \sep code
%% or \MSC[2008] code \sep code (2000 is the default)
\MSC 0000 \sep 1111
\end{keyword}

\end{frontmatter}

%% \linenumbers

%% main text

\include{Introduction}
\include{Methodology}
\include{Discretization}
\include{Results}

\include{Conclusions}

%% The Appendices part is started with the command \appendix;
%% appendix sections are then done as normal sections
\appendix
\include{Appendix}

%% If you have bibdatabase file and want bibtex to generate the
%% bibitems, please use
%%
 % \nocite{*}
 \bibliographystyle{elsarticle-num} 
 \bibliography{main}

%% else use the following coding to input the bibitems directly in the
%% TeX file.

% \begin{thebibliography}{00}

% %% \bibitem{label}
% %% Text of bibliographic item

% \bibitem{}

% \end{thebibliography}
\end{document}

%% file: Introduction.tex
\section{Introduction}

\label{sec:introduction}

%\sd{commento1} 
%\vr{commento2} 
%\vf{commento3}
%\pa{commento4}
%\cdf{commento5}

The Poisson-Boltzmann equation (PBE) is a cornerstone of electrostatic modeling in soft matter physics, biophysics, and colloidal science \cite{JPCBSI,BesleyReview}. It plays a crucial role in understanding how charged particles interact in various environments, including biological systems, colloidal suspensions, and electrolytic solutions. At its core, the PBE provides a mathematical framework to describe the mean field electrostatic potential generated by a distribution of fixed charges in a dielectric medium immersed in another one. The latter is usually a more polarizable dielectric medium representing the solvent and containing mobile ions, described as potential-dependent distributions \cite{Blossey2023}.

In biological systems, the PBE is a crucial instrument for modeling interactions at the molecular level, such as protein-protein binding, nucleic acid stability, and enzyme activity. It helps predict how biomolecules behave in different ionic environments, a critical factor in drug design, molecular recognition, and understanding the stability of macromolecular structures. Beyond biology, the PBE is equally essential in materials science, where it is used to predict the behavior of colloidal particles, understand self-assembly processes, and design new materials with specific electrostatic properties.

The PBE is inherently nonlinear, making it challenging to solve analytically or even numerically. For many practical applications, however, the linearized version of the PBE (LPBE) is used, especially when the system involves low surface potentials. However, even the LPBE requires sophisticated mathematical techniques and numerical methods for its accurate resolution, particularly in complex geometries or large-scale systems. Indeed, the model used to describe the molecular surface (MS) has great impact on the potential resulting from the solution of the equation \cite{Between,NanoShaper}, and it has been shown that subtle, but still biologically relevant, effects descend from the intricate geometries of the MS~\cite{Honig2009}.

Advances in computational methods have made it possible to tackle the PBE with greater precision \cite{MIBPB2}, allowing for more detailed and accurate modeling of electrostatic interactions in various systems. These developments have led to new insights into phenomena like ion binding, charge screening, and the stabilization of charged interfaces, further solidifying the PBE as an essential tool in the study of electrostatics.

%In summary, the Poisson-Boltzmann equation is fundamental to the understanding and prediction of electrostatic phenomena in many scientific fields. 
Present challenges for this approach are its ability to describe large systems (as those observed by Cryo-EM experimental techniques), its accuracy in matching experimental energies and local electric fields~\cite{iwahara2023direct,yu2022measuring}, its computational cost, an essential requirement for its highly repetitive execution required by pKa states description \cite{MCCE2} or Constant-pH Molecular Dynamics simulations~\cite{ConstPh}. These considerations fuel the quest for more efficient resolution techniques, which will strengthen and expand their applicability and offer more profound insights into the behavior of charged systems at both the macroscopic and microscopic levels.

In this work, we present several advances in the model and in the resolution of the LPBE, where the contribution of analytical calculations is instrumental in improving the solution's accuracy without increasing the computational cost. 
Namely, we use the analytical information on the MS, as provided by the NanoShaper software \cite{NanoShaper,NanoShaper-vmd} to enable a local Taylor expansion for the electrostatic potential at the MS, which in turn is used to improve the discretization of the equation. By suitable use of Green's identities, we derive an expression for the interaction between the solvent counterions and the charges of the solute, which avoids integrating over the entire solvent volume. Finally, we propose to use a variable-resolution grid to apply accurate boundary conditions (BCs) at a limited computational cost. 
To validate the improved accuracy of the method, we compare with analytical expressions for a system of many charged, non-overlapping, dielectric spheres immersed in a Debye-H{\"u}ckel solvent. Next, we apply the method to systems of biological interest, namely proteins and nucleic acids. 
The proposed advances are implemented in NextGenPB, a Finite Element Method PB solver, which is made available at \url{https://github.com/concept-lab/NextGenPB} or \url{https://concept.iit.it/downloads}.
In the following sections, we discuss the enhanced solution method, present experimental results and provide a final discussion and conclusions.

%% file: Methodology.tex
\section{Physical models and Mathematical foundations}

\subsection{Poisson-Boltzmann equation}

Consider a bounded domain $ D = \Omega_m \cup \Omega_s \cup \Gamma $, where $\Omega_m$ represents the region of space occupied by the solute (usually a molecule), $\Omega_s$  is the solvent region, and $\Gamma$ is the closed surface separating them. In the context of biomolecular electrostatics calculations, a molecule is represented as a union of possibly overlapping linear dielectric balls, the atoms, immersed in a high-dielectric solvent. The relative dielectric constant of the solute, $\varepsilon_{r,m}$ usually ranges between 2 and 4, reflecting the electronic polarization of the molecule. The solute contains all the explicitly assigned partial charges centered on atoms. In contrast, $\varepsilon_{r,s} $, i.e., the solvent's relative dielectric constant, is higher, nearly 80, accounting for the presence of mobile ions that respond to the local electrostatic potential. From the physical point of view, both atomic charges and mobile ions constitute the free charge. Still, there is a significant modelistic distinction between them since the former is assigned as an input to the equation. At the same time, the ionic density results from the calculation since the model prescribes a relationship between the ionic response and the local electric field.

In this context, the polarization response is described by a piece-wise constant dielectric function:
\begin{equation}
    \varepsilon(\mathbf{r})=
    \left\{
    \begin{array}{ll}
        \varepsilon_m= \varepsilon_0\varepsilon_{r,m} & \text{if } \mathbf{r}\in \Omega_m\\
        \varepsilon_s= \varepsilon_0\varepsilon_{r,s} & \text{if } \mathbf{r}\in \Omega_s
    \end{array}
    \right. \; ,
\end{equation}
where $\varepsilon_0$ is the vacuum permittivity, and $\varepsilon_m$ and $\varepsilon_s$ are the permittivities of molecule and solvent, respectively. 
Gauss's law can describe the electrostatics of this system for the electric displacement in differential form:
\begin{equation}
    \nabla \cdot \left( \varepsilon(\mathbf{r})\nabla\phi \right) =
         - \rho^{free} = - \rho^f- \rho^s 
         %\approx - \rho^f + c(\mathbf{r})\phi 
         \quad \mathbf{r}\in D \;, 
\label{eq:LPB_operator}
\end{equation}
where $\rho^f $ is the fixed charge density present only inside the molecule, and $\rho^s $ is the ionic charge present only in the solvent.
The fixed charge density, composed of point charges, is defined as:
\begin{equation}
    \rho^f(\mathbf{r}) =  \sum_{i =1}^N q_i \delta(\mathbf{r}-\mathbf{r}_i)
\end{equation}
where $\mathbf{r}_i$ is the position of the $i-$charge. The concentrations of the different ionic species inside the solvent can be described by assuming that the charge carriers are immaterial, that only feel the local electrostatic potential and the thermal bath and that they obey the overall electroneutrality condition: $\sum_{j=1}^J {z_j e\rho_j^b} = 0$, where  $J$ is the number of ionic species, $e$ is the charge of a proton, and $z_j$ and $\rho_j^b$ denote the valence and bulk concentration of ions of species $j$, respectively.  Under these assumptions, one can impose that, at thermal equilibrium, the local concentration of each species is related to that of the same species in the bulk by a Boltzmann weight. This leads to the following expression for the charge density in the solvent:
\begin{equation}
    \rho^s(\mathbf{r}) =  \sum_{j =1}^J ez_j \rho_j^b\exp[-\dfrac{e z_j\phi(\mathbf{r})}{K_b T}] \chi_{\Omega_s}(\mathbf{r})
\end{equation}
($K_b$ and $T$ are Boltzmann's constant and absolute temperature, respectively). 

Since the ions are only located in the solvent, the characteristic function $\chi_{\Omega_s}$ equals 1 when $\mathbf{r}\in\Omega_s$ and 0 elsewhere. 
This yields the Poisson-Boltzmann equation:
\begin{equation}
    \nabla \cdot \left( \varepsilon(\mathbf{r})\nabla\phi \right) =
    -\sum_{i =1}^N q_i \delta(\mathbf{r}-\mathbf{r}_i)
    -\sum_{j =1}^J \left[ ez_j \rho_j^b\exp{-\dfrac{e z_j\phi(\mathbf{r})}{K_b T}}\right] \chi_{\Omega_s}(\mathbf{r}) \;
\end{equation}
In case the Stern layer model is used, $\chi_{\Omega_s}=0$ also inside it. 
For low-charged systems, the resulting potential in solution may satisfy the following relationship: $|\frac{e z_j\phi(\mathbf{r})}{K_b T}| \ll 1$. Under this and electroneutrality assumptions, the local concentration of ions in the solvent can be approximated as a linear function of the potential:
\begin{equation}
    \rho^s(\mathbf{r}) \approx  - \varepsilon_s k_D^2 \chi_{\Omega_s}(\mathbf{r}) \phi(\mathbf{r}) =  - c(\mathbf{r})\phi(\mathbf{r}) \;,
\end{equation}
where $k_D = \sqrt{\frac{e^2 }{\varepsilon_s K_b T}\sum_{j=1}^J z_j^2 \rho_j^b}$ is the Debye factor for the solution, reciprocal of the so-called Debye screening length. This approximation leads to the Debye-H{\"u}ckel (D-H), or linear Poisson-Boltzmann equation (LPBE):

\begin{equation}
    \nabla \cdot \left( \varepsilon(\mathbf{r})\nabla\phi \right) =
    -\sum_{i =1}^N q_i \delta(\mathbf{r}-\mathbf{r}_i)
    + c(\mathbf{r})\phi(\mathbf{r}) \qquad \mathbf{r}\in D \;. 
    \label{eq:LPBE}
\end{equation}
Equation \eqref{eq:LPBE} is a linear, elliptic partial differential equation. It is more manageable numerically and analytically than its nonlinear counterpart, making it the most commonly used equation for studying the electrostatics of biomolecular systems.

\paragraph{Electrostatic energy}
\label{par:electrostatic_energy}
The derivation of the energy of a system described by linear Poisson-Boltzmann model with $N$ fixed point charges in the solute can be found, for instance, in  \cite{sharp1990calculating} and it takes the form:
\begin{equation}
    E = \dfrac{1}{2}\int_{\Omega} \rho^f(\Tilde{\mathbf{r}}) \phi(\Tilde{\mathbf{r}}) \mathrm{d}V  
      = \dfrac{1}{2} \sum_{i =1}^N q_i  \phi(\mathbf{r}_i) \; ,
\label{eq:sharp_energy}
\end{equation}
where $\phi$ is the total potential, excluding the so-called self-potential, which is the one generated by $q_i$ at $\mathbf{r}_i$. $E$ can be considered the electrostatic component of a solvated system's free energy, where the solvent's degrees of freedom have been averaged out. A convenient way to represent the electrostatic energy \eqref{eq:sharp_energy} is given in work \cite{rocchia2001extending}, where the total potential is partitioned in three contributions: the Coulombic term $\phi_{coul}$, which can be calculated analytically, the $\phi_{pol}$ term, due to the polarization charges arising at the MS where the dielectric constant varies, and $\phi_{ion}$, the potential generated by the counterions in the solvent. Both $\phi_{coul}$ and $\phi_{ion}$ account for the screening due to the continuum medium in which their sources are located.

\begin{equation}
    E = \frac{1}{2}\sum_{i=1}^N q_i \left( \phi_{coul}(\mathbf{r}_i) + \phi_{pol}(\mathbf{r}_i) + \phi_{ion}(\mathbf{r}_i) \right)\, .
    \label{eq:energyPart}
\end{equation}
This partitioning of the potential and, consequently, of the energy allows for the analytical estimation of the Coulombic term, avoiding the self-energy, and for the calculation of the polarization term. 
As per the ionic term, its source is the local ionic imbalance, which extends over the entire solvent until where the potential vanishes. Since volume integration of this latter term, which turns out to be overall smaller than the previous ones, is prohibitive, it is usually avoided. Alternatively, the ionic energy contribution can be calculated by taking the difference of the so-called \emph{grid energy} in two successive runs at different ionic strengths, namely $0$ and the actual one. Grid energy is half the sum of the fixed charges mapped onto the grid points times the numeric potential evaluated at the corresponding points.\\

\subsection{Electrostatic potential partitioning}
\label{sect.Partitioning}
In this section, we use two expressions for the electrostatic potential partitioning based on Green's identities, one when the potential is evaluated inside the solute and the other when it is evaluated in the solvent. They are derived in \ref{appendix_internal_potential} and \ref{appendix_external_potential2}. These formulations are convenient for practical energy calculation and for assessing the quality of existing boundary conditions. 
\\
Let's consider the potential evaluated at a point $\mathbf{r}$ belonging to a volume region $\Omega_m$ enclosed in a surface $\Gamma$. This region, that in the PB case represents the solute, is surrounded by the electrolytic solution, and contains the free charge distribution $\rho^f$.
As shown in \ref{appendix_internal_potential}, this potential can be expressed as follows $\forall \mathbf{r}\in\Omega_m$:
\begin{equation}
\begin{split}
    \phi(\mathbf{r}) = \int_{\Omega_m} \dfrac{\rho^f}{4\pi\varepsilon_m\|\Tilde{\mathbf{r}} -\mathbf{r}\|} \dd V
    -\int_{\Gamma} \dfrac{\mathbf{D}(\Tilde{\mathbf{r}}) \cdot \mathbf{n}(\Tilde{\mathbf{r}})}{4\pi \varepsilon_m \| \Tilde{\mathbf{r}}-\mathbf{r} \|}  \dd S + & 
    \int_{\Gamma} \phi(\Tilde{\mathbf{r}}) \dfrac{(\Tilde{\mathbf{r}}-\mathbf{r}) \cdot \mathbf{n}(\Tilde{\mathbf{r}})}{4\pi \| \Tilde{\mathbf{r}}-\mathbf{r} \|^3}  \dd S \, .  
\end{split}
\label{eq:pot_in_partitioning}
\end{equation}
\noindent
The first term in the RHS is $\phi_{coul}$, that is the Coulombic part of the potential.\\ 
In agreement with the continuum electrostatic model, the surface polarization charge is distributed on $\Gamma$, and it is equal to:
\begin{equation}
    \sigma_{pol}(\mathbf{r}) = -\left (\mathbf{P}_2(\mathbf{r}) -\mathbf{P}_1(\mathbf{r})\right ) \cdot \mathbf{n}_{21}(\mathbf{r}) \quad \mathbf{r}\in \Gamma\; ,
\end{equation}
where $\mathbf{n}_{21}$ is the unit normal pointing from the molecule to the solvent, and $\mathbf{P}_1(\mathbf{r})$ and $\mathbf{P}_2(\mathbf{r})$ are the polarization vectors in the molecule and in the solvent respectively, evaluated at $\mathbf{r}\in\Gamma$. Therefore, the surface polarization charge in this system~is:
\begin{equation}
    \sigma_{pol}(\mathbf{r}) = \varepsilon_0 \left(\dfrac{1}{\varepsilon_s} - \dfrac{1}{\varepsilon_m}\right)\mathbf{D}(\mathbf{r})\cdot\mathbf{n}(\mathbf{r}) \qquad \mathbf{r}\in \Gamma  \; ,
\label{eq:pol_distr}
\end{equation}
and the corresponding potential is:
\begin{equation}
\label{eq:phi_pol}
\begin{split}
    \phi_{pol} (\mathbf{r}) &= \int_{\Gamma} \dfrac{\sigma_{pol}(\Tilde{\mathbf{r}})}{4\pi\varepsilon_0 \|\Tilde{\mathbf{r}} -\mathbf{r}\|} \dd S =
     \left(\dfrac{1}{\varepsilon_s} - \dfrac{1}{\varepsilon_m}\right)\int_{\Gamma} \dfrac{\mathbf{D}(\Tilde{\mathbf{r}}) \cdot \mathbf{n}(\Tilde{\mathbf{r}})}{4\pi \| \Tilde{\mathbf{r}}-\mathbf{r} \|}  \dd S \; .
\end{split}
\end{equation}
Considering this term, we can rewrite the potential in Eq.~\eqref{eq:pot_in_partitioning} to make the polarization term more explicit:
\begin{equation}
\begin{split}
    \phi(\mathbf{r}) = &\phi_{coul} (\mathbf{r}) 
                 + \left(\dfrac{1}{\varepsilon_s} - \dfrac{1}{\varepsilon_m}\right)\int_{\Gamma} \dfrac{\mathbf{D}(\Tilde{\mathbf{r}}) \cdot \mathbf{n}(\Tilde{\mathbf{r}})}{4\pi \| \Tilde{\mathbf{r}}-\mathbf{r} \|}  \dd S  \\
               &+ \int_{\Gamma} \phi(\Tilde{\mathbf{r}}) \dfrac{(\Tilde{\mathbf{r}}-\mathbf{r}) \cdot \mathbf{n}(\Tilde{\mathbf{r}})}{4\pi \| \Tilde{\mathbf{r}}-\mathbf{r} \|^3}  \dd S  - \dfrac{1}{\varepsilon_s}\int_{\Gamma} \dfrac{\mathbf{D}(\Tilde{\mathbf{r}}) \cdot \mathbf{n}(\Tilde{\mathbf{r}})}{4\pi \| \Tilde{\mathbf{r}}-\mathbf{r} \|}  \dd S 
\end{split}
\qquad \forall \mathbf{r}\in\Omega_m \; .
\label{eq:pot_inside_final}
\end{equation}
When comparing Eq.~\eqref{eq:pot_inside_final} with \eqref{eq:energyPart}, one can immediately identify the sum of the last two terms as the potential due to the ionic charge density in the solution. This expression, explicitly derived in \ref{appendix_external_potential1}, converts a volumetric term extending over a possibly huge domain into a much more localized surface term. This equivalence is particularly convenient for calculating the ionic direct energy term in just one solving process, in contrast, for instance, to what is described in Sect.~\ref{par:electrostatic_energy}, as it will be shown in Sect.~\ref{sect:ionicCalc}.\\

One could repeat a similar analysis employing Green's identities when considering the potential in the solvent region. Following the derivation performed in \ref{appendix_external_potential2}, this leads to the following formulation for the electrostatic potential outside the molecule $\forall \mathbf{r}\in\Omega_s$:
\begin{equation}
\begin{split}
\phi(\mathbf{r}) = {}& 
    \int_{Stern~layer} \phi(\Tilde{\mathbf{r}})\dfrac{k_D^2 e^{-k_D\| \Tilde{\mathbf{r}}-\mathbf{r} \|}}{4\pi \| \Tilde{\mathbf{r}}-\mathbf{r} \|} \dd V
    + \int_{\Gamma} \dfrac{e^{-k_D\| \Tilde{\mathbf{r}}-\mathbf{r} \|}\mathbf{D}(\Tilde{\mathbf{r}}) \cdot \mathbf{n}(\Tilde{\mathbf{r}})}{4\pi \varepsilon_s \| \Tilde{\mathbf{r}}-\mathbf{r} \|} \dd S  +\\
    &-\int_{\Gamma} \phi(\Tilde{\mathbf{r}}) \dfrac{e^{-k_D\| \Tilde{\mathbf{r}}-\mathbf{r} \|}(1+k_D\| \Tilde{\mathbf{r}}-\mathbf{r} \|)(\Tilde{\mathbf{r}}-\mathbf{r}) \cdot \mathbf{n}(\Tilde{\mathbf{r}})}{4\pi\| \Tilde{\mathbf{r}}-\mathbf{r} \|^3}  \dd S \; .
\end{split}
\label{eq:out_pot}
\end{equation}

As customary in continuum electrostatics, this can be interpreted as the sum of the potential generated by a surface charge distributed on the MS, of that generated by a layer of ideal dipoles located also on the MS, plus a contribution spread over the Stern layer. The electrolytic solution screens both sources of potential. This expression will be used in Sec.~\ref{sec:BCs} to assess the quality of existing boundary conditions.

\subsection{Boundary conditions}
\label{sec:BCs}
Choosing convenient boundary conditions is essential in setting up the PBE solution.
Physical considerations tell us that the electrostatic potential and field are expected to approach zero at a sufficient distance from the region where the fixed charges are located.  Consequently, suitable boundary conditions could be the homogeneous Dirichlet ones, where the potential is set to zero at the boundary of the solution domain. However, this can be acceptable only if the \emph{perfil}, that is the ratio between the size of the solute and that of the computational domain, is relatively small. 
To allow a larger \emph{perfil}, analytical approximations for the potential at an intermediate distance from the solute are often used. For example, the so-called Coulombic or, better, Debye-Hückel boundary conditions. If one considers a system with $N$ fixed charges and a solvent containing a dissociated monovalent salt, the D-H solution takes the form:
\begin{equation}
    \phi(\mathbf{r}) = \sum_{i=1}^N \frac{q_ie^{-k_D\|\Tilde{\mathbf{r}} -\mathbf{r}\|}}{4 \pi \varepsilon_s \|\Tilde{\mathbf{r}} -\mathbf{r}\|}\; .
\end{equation}
There are two main differences between the potential generated by this model, which has been called the "fully penetrating solvent" model~\cite{RocchiaMCM}, and a more realistic one, where the charges are not directly exposed to the solvent but are instead embedded into a low dielectric solute. One is that the approximated potential misses the dipolar contribution originated in the low-polarization medium. The other is that counterions in the approximated model occupy the region that in reality is occupied by the low dielectric medium. This second term is more impactful than the first, which decays as a screened dipolar potential does.
Suppose we consider the simplest among the few known analytical solutions for the LPBE, namely the spherical case. In that case, we can easily derive the exact potential outside a sphere of the radius $R$ with a single charge $q$ located in its center:
\begin{equation}
    \phi(r) = \dfrac{q}{4\pi \varepsilon_s r(1+k_\text{$D$} R)}e^{-k_\text{$D$}(r-R)}  = \phi_{coul}(r) \dfrac{e^{k_\text{$D$} R}}{(1+k_\text{$D$} R)}
\end{equation}
It can be easily seen that the difference between the actual potential and its D-H approximation tends to zero of the same order as the actual potential does, so the D-H approximation cannot be considered an actual asymptotic solution.

Let's now consider the most general case of Eq. \eqref{eq:out_pot}. We observe that the first two terms are prevalent at larger distances than the second one, since the potential generated by a dipole layer decays faster than that of a surface charge. One can thus conclude, at least qualitatively, that a more accurate asymptotic solution should be written as the sum of Debye-H{\"u}ckel-like terms centered on points lying on the MS or in the Stern layer. This consideration, however, does not provide details on the parameters of these terms, since their knowledge requires that of the electric displacement at the MS, which become available only after the PBE is solved.
Along the same qualitative lines, one could expect that the discrepancy between the actual potential and that obtained by the penetrating solvent approximation should increase with the volume of the solute, which excludes more counterions, and with its absolute net charge, which reduces the possibility of cancellations in the ionic reaction.

%of As the distance between the evaluation point and the surface increases, the second term of Eq. \eqref{eq:out_pot} decays faster than the first term. Thus, we can approximate the expression of the potential at the domain boundary as follows:
%\begin{equation}
%    \phi(\mathbf{r}) \approx
%    \int_{\Gamma} \dfrac{\mathbf{D}(\Tilde{\mathbf{r}}) \cdot \mathbf{n}(\Tilde{\mathbf{r}})}{4\pi \varepsilon_s \| \Tilde{\mathbf{r}}-\mathbf{r} \|}e^{-k\|\Tilde{\mathbf{r}} -\mathbf{r}\|}  \dd S \; .
    %\label{eq:phi_apprax_ext}
%\end{equation}
% \\ 

\section{The numerical model}

\label{sect:primal_mixed}

In devising a convenient discretization strategy for the model described above, we have been guided by two main objectives: on the one hand, we want to achieve the best possible accuracy
for the electrostatic potential and energy calculation near the molecular surface where the discontinuity occurs; on the other hand, the extreme size of the problems typically 
being considered demands for a discretization scheme generating matrices with simple structure and small stencil. To attain the former goal, we adopt special quadrature formulas based on the physically motivated averaging of the electrical permittivity introduced in Sect.~\ref{sec:eps_disc}.
To achieve the latter goal, we limit the complexity of the discretization stencils by working with low order Finite Element basis functions on a Cartesian grid, and applying a reduced integration~\cite{hughes_reduced_integration} approach.
%of the stiffness matrix, 
Our discretization method can be represented as a mixed formulation of the elliptic PDE at hand, 
{\it i.e.,} we start by reformulating the second order differential problem as a system
of two first order problems and then we choose suitable function spaces to represent 
both the \emph{primal variable} (the electrostatic potential $\phi$) and the \emph{dual variable} (the electric displacement vector $\mathbf{D}$).
In particular, we use the primal--mixed formulation~\cite{ROBERTS1991523,robey_pm}, whereby global continuity of $\phi$ is enforced, while a partially discontinuous representation of $\mathbf{D}$ is tolerated. 
%The discontinuity of $\mathbf D$ allows to locally
We then eliminate the degrees of freedom related to the dual variable, a procedure commonly referred to as \emph{static condensation}~\cite{static_condensation}. 
Hierarchical (de)refinement also allows for a significant reduction of the problem
unknowns and is a crucial feature of our implementation. 
The approach used to accommodate for non-conforming meshes with so-called \emph{hanging nodes} is described in some detail in previous work~\cite{bimpp,landslide} therefore, in Sect.~\ref{sect:pmfem}
we assume that, for sake of brevity and readability, our discretization domain consists of a full tensor-product Cartesian grid.

\subsection{Primal--Mixed Finite Elements Discretization of the LPBE}\label{sect:pmfem}
%
%The Poisson-Boltzmann equation is a second-order elliptic PDE of the form:
%\begin{equation}
 %   \nabla \cdot \left( \varepsilon(\mathbf{r})\nabla\phi \right) =
 %        - \rho \qquad  \mathrm{on} \; \Omega \subset \mathbb{R}^3 
%\label{eq:pbe0}
%\end{equation}
%
As already mentioned above, we start the derivation by formulating
 LPBE \eqref{eq:LPBE}, over a domain $ D \subset \mathbb{R}^3$ as a system of two first order equations:
\begin{equation}
    \begin{cases}
    \dfrac{1}{\varepsilon}\mathbf{D} + \nabla\phi &= 0   \\
    \nabla\cdot{\mathbf{D}} + c\ \phi & = \rho^f.
\end{cases}
\label{eq:pbe1}
\end{equation}
We assume for the time being that homogeneous  Dirichlet--type Boundary conditions are to be enforced on the whole boundary $\partial D $ of $D$ , but a more in-depth
discussion of BCs will be dealt with in Sect.~\ref{sec:BCs}.

In order to numerically solve~\eqref{eq:pbe1} via a  Galerkin/Finite Element approximation we must, first of all, state the problem in weak  form. To this end, we multiply both equations in~\eqref{eq:pbe1} by suitable \emph{test functions} and perform
integration by parts to obtain:

find $(\mathbf{D},\phi)\in U\times  W$ such that $\forall (\pmb{\sigma},v)\in U\times W$:
\begin{equation}
\begin{aligned}
    &\int_{D}\frac{1}{\varepsilon}\ \mathbf{D}\cdot\pmb{\sigma} \, \dd V + 
    \int_{D} \nabla\phi \cdot\pmb{\sigma} \, \dd V  
    && = && 0 \, ,\\
    &\int_{D}\mathbf{D}\cdot \nabla v \, \dd V -
    \int_{D} c\ \phi \ v \, \dd V&&
    = && - \int_{D} \rho^f\ v \, \dd V \, ,
\end{aligned}
\label{eq:pbe2'}
\end{equation}
where $U\equiv \left( L^2(D) \right)^3$ and $W \equiv H^1_0(D)$.
Introducing the bilinear forms $a(\cdot,\cdot): U\cross U \rightarrow \mathbb{R}$ and $b(\cdot,\cdot): U\cross W \rightarrow \mathbb{R}$, 
$c(\cdot,\cdot): W\cross W \rightarrow \mathbb{R}$
and the linear operator $R:W \rightarrow \mathbb R$,
    
\begin{equation}
\begin{aligned}
    a( \mathbf{u}, \mathbf{z}) &= \int_{D}\frac{1}{\varepsilon} \mathbf{u}\cdot\mathbf{z} \,\dd V \, ,\\
    b ( \mathbf{u}, v) &= \int_{D} \mathbf{u} \cdot \nabla v \, \dd V \, ,\\
    \kappa (w, v) &= \int_{D} c\ w\ v \, \dd V \, ,\\
    R(v) &= \int_{D} \rho^f\ v \, \dd V \label{RVop_} \, .
\end{aligned}
\end{equation}
Eq. \eqref{eq:pbe2'} becomes:
\begin{equation}\begin{split}
&\mathrm{find}\; (\mathbf{D},\phi)\in U\cross W\; s.t.\\
&\begin{cases}
    a(\mathbf{D},\pmb{\sigma}) + b (\pmb{\sigma}, \phi) =0\\
    b(\mathbf{D},v)  -\kappa(\phi, v) = -R(v)
\end{cases}\\
&\forall (\pmb{\sigma},v) \in U\cross W \: 
\label{eq:pbe2''}
\end{split}
\end{equation}

The Galerkin/Finite Element method for the numerical approximation of \eqref{eq:pbe2''} consists of looking for an approximate solution  $(\mathbf{D}_h,\phi_h) \in U_h\cross W_h$, where 
\begin{equation*}
\begin{split}
    &W_h \subset W, \;\; \dim(W_h) = N^h_v < \infty\\
    &U_h \subset U, \;\; \dim(U_h) = N^h_e <\infty
\end{split}
\end{equation*}
are two families of \emph{finite dimensional} subspaces  depending on the parametrized by the \emph{mesh size} $h$ \emph{s.t.}
$U_h\rightarrow U$ and $W_h\rightarrow W$ for $h\rightarrow 0$.
%where $N^h_e$ is the number of edges and $N_v^h$ is the number of vertices. 
Then, the discretized version of \eqref{eq:pbe2''} takes the form:
\begin{equation}\label{eq:pbe2'''}
\begin{split}
&\mathrm{find}\; (\mathbf{D}_h,\phi_h)\in U_h\cross W_h\; s.t.\\
&\begin{cases}
    a(\mathbf{D}_h,\pmb{\sigma}) +  b (\pmb{\sigma}, \phi_h) =0 \\
    b(\mathbf{D}_h,v) -\kappa(\phi_h, v)  = -R(v) 
\end{cases}\\
&\forall (\pmb{\sigma},v)\in U_h\cross W_h   
\end{split}
\end{equation}
In order to guarantee the \emph stability of~\eqref{eq:pbe2'''}, the
chosen pair of finite dimensional spaces $U_h$ and $W_h$ must satisfy the Ladyzhenskaya–Babuška–Brezzi (LBB) condition~\cite{Boffi2013,ROBERTS1991523} which is simply attained if $\nabla W_h \subset U_h$~\cite{robey_pm}.
To enforce the latter condition we chose as $W_h$ the set of piecewise tri--linear continuous functions on the triangulation, while we let $U_h$ be comprised of lowest-order  N\'ed\'elec edge elements of the first kind~\cite{nedelec1980mixed,nedelec1986new}.
Denoting by $\{ \pmb{\sigma}_i \}$ a basis of $U_h$ and by $\{w_k\}$ a basis of $W_h$, we can write  $\mathbf{D}_h$ and $\phi_h$ as an appropriate linear combination of these bases,~i.e.
\begin{equation}
    \mathbf{D}_h = \sum_{j=1}^{N_e^h} D^h_j\pmb{\sigma}_j \; , \quad \phi_h = \sum_{k=1}^{N_v^h} \phi^h_k w_k \; .
\end{equation}
We note that the dimension $N^h_e$ of $U_h$ is the number of edges in the mesh and dimension $N_v^h$ of $W_h$ is the number of interior vertices.
We shall then request:
\begin{equation}
\begin{cases}
    \sum_j D^h_j a(\pmb{\sigma}_j,\pmb{\sigma}_i) + \sum_k \phi^h_k b (\pmb{\sigma}_i, w_k) = 0 &\forall i \, ,\\
    \sum_j D^h_jb(\pmb{\sigma}_j,w_n)  - 
    \sum_k
    \phi^h_k \kappa (w_k, w_n) = -R(w_n)  &\forall n .
\end{cases}
\label{eq:pbe3}
\end{equation}
The matrix form of \eqref{eq:pbe3} is
\begin{equation}
\begin{bmatrix}
    \Tilde{A} & B^T\\
    B & -C
\end{bmatrix}
    \left[\begin{array}{l} \mathbf{D}^h\\ \, \pmb{\phi}^h \end{array} \right]
    =
    \left[\begin{array}{l} 0\\ -\pmb R \end{array} \right] \; ,
\label{eq:pbe3'}
\end{equation}
where $\Tilde{A} = [a(\pmb{\sigma}_i,\pmb{\sigma}_j)] = [a_{i,j}] $, 
$B^T = [b(\pmb{\sigma}_i, w_k)] =[b_{i,k}]$, $\mathbf{D}^h=[D^h_1,\dots,D^h_{N^h_e}]$, 
$\pmb{\phi}^h = [\phi^h_1,\dots \phi^h_{N^h_v}]$, 
$C = \left[ \kappa(w_k, w_n) \right] = [c_{k,n}]$ and
$\pmb R = [R(w_1),\dots, R(w_{N^h_v})]$.
\\
As anticipated in the introduction to Sect.~\ref{sect:primal_mixed}, the structure 
and bandwidth of the matrices in~\eqref{eq:pbe3'} can be highly simplified by approximating the integrals in~\eqref{eq:pbe2'} by means of low order quadratures.
In particular, it is easily shown that, by selecting the cell vertices as quadrature nodes, $\Tilde{A}$, ends up being a diagonal, and therefore easily invertible, matrix.
We can take advantage of the simple form of $\tilde{A}$ to reduce the total number
of unknowns in the systems via a Schur--complement approach (often referred to as \emph{static condensation} in the jargon of Finite Elements practitioners~\cite{zienkiewicz1977finite}).
Indeed, solving the first row of~\eqref{eq:pbe3'} for $\mathbf{D}^h$ we get
$$
\mathbf{D}^h = - \Tilde{A}^{-1} B^T \boldsymbol{\phi}^h 
$$
and plugging the latter into the second row
\begin{equation}\label{eq:matrixequation}
 \left[- B \left( \Tilde{A}^{-1} B^T \right) - C \right] \boldsymbol{\phi}^h
= A\ \pmb{\phi}^h = - \pmb{R}.   
\end{equation}
As for the choice of the quadrature rule, the obvious choice for a rule with nodes at
the vertices is that of the trapezoidal rule; it is easily shown that upon making
such choice, the matrices in~\eqref{eq:matrixequation} correspond to those that
would be obtained by taking the \emph{harmonic average}, that is the average of the inverses, of the permittivity over the
edges and applying a standard Finite Element discretization of the problem~\cite{xu1999monotone,markowich1988inverse}.
In the next sections, we will show how a different choice of the quadrature weights (or equivalently a choice of the averaging coefficients for the permittivity on the edges)
can significantly improve the performance of the discretization near the molecular surface.

\subsection{Coefficient Discontinuity: equation discretization across the molecular surface $\Gamma$}
\label{sec:eps_disc}
As described, for the solution of the LPBE, we opt for an orthogonal cubic finite elements method (FEM). In this scheme, we represent our potential in a first-order Lagrangian basis. According to this description, the potential and similar quantities defined on the nodes are tri-linearly interpolated in off-grid positions. It descends from this approach that the interpolated potential changes are always linear along grid edges, and correspondingly, the electric field is constant on them. This reconstruction scheme perfectly fits the continuum electrostatics model when a grid edge is entirely inside the same medium (solute or solvent). However, it conflicts with the physical model when an edge crosses the surface, separating two different media. Indeed, the continuum electrostatics theory prescribes that the potential is continuous across the jump in dielectric constant, likewise the orthogonal component of the dielectric displacement. This causes a step discontinuity on the electric field's orthogonal components, which a linear interpolation along adjacent nodes cannot recover.
Conventional solutions to this problem can be increasing the grid resolution, possibly locally, using irregular grids to localize the nodes at the discontinuity or changing the shape of the elements so that they better fit that of the MS. All of these solutions, however, entail some increase in the degree of complexity of the method. 
In our quest for increased accuracy at low computational impact, we note that the information that we can get from NanoShaper \cite{NanoShaper} is richer than simply telling whether a point is inside or outside the volume enclosed by the MS. We have analytical information related to where exactly the edges intersect the MS, together with the corresponding normal. We aim to use this information to improve how the equation is discretized on the MS-crossing edges.
\\
Let's consider two adjacent nodes, \textit{1} and \textit{2} along a given MS-crossing edge and let's call \textit{0} the point where the edge intersects the MS. Let's then consider the two first-order Taylor expansions of the potential centered at the point of intersection along that edge. One expansion holds in medium \textit{1}, while the other in medium \textit{2}:
 
\begin{equation}
    \phi(\mathbf{r}_1) = \phi (\mathbf{r}_0^-) - {E}_{\nu}(\mathbf{r})\Bigr|_{\mathbf{r}_0^-}  ({r}_1^{\nu} -{r}_0^{\nu,-} ) + O(\|\mathbf{r}_1-\mathbf{r}_0\|^2),
    \label{eq:taylor1}
\end{equation}
\begin{equation}
    \label{eq:taylor2}
    \phi(\mathbf{r}_2) = \phi (\mathbf{r}_0^+) - {E}_{\nu}(\mathbf{r})\Bigr|_{\mathbf{r}_0^+} ({r}_2^{\nu} -{r}_0^{\nu,+} ) + O(\|\mathbf{r}_2-\mathbf{r}_0\|^2), 
\end{equation}
where $\mathbf{r}_0$ is the exact position of the intersection, $\mathbf{r}_1$ and $\mathbf{r}_2$ are the positions of the nodes, $\nu$ is the coordinate direction of the edge and $E_{\nu}$ is the $\nu$-component of the electric field. The symbols $\mathbf{r}_0^-$ and $\mathbf{r}_0^+$ denote the position's limit values as they approach the surface from the two different media and coincide. \\
The electric field  at the intersection point can be written as the sum of the vector components normal and tangential to the MS.
The electric field at the MS is not continuous, but it can be conveniently represented as the sum of its tangential component vector, which is continuous, and of the normal component vector of the electric displacement, also continuous, divided by the local dielectric constant: 
$\mathbf{E}(\mathbf{r}_0^{\pm}) = \mathbf{E_t}(\mathbf{r}_0)+ \dfrac{\mathbf{D_n}(\mathbf{r}_0)}{\varepsilon(\mathbf{r}_0^{\pm})}$.\\
\noindent
By denoting with $\alpha$ the edge fraction in medium \textit{1}, we obtain ${r}_1^{\nu} -{r}_0^{\nu,-} = - h_{\nu}\alpha$ and ${r}_2^{\nu} -{r}_0^{\nu,+} = h_{\nu}(1-\alpha)$, where $h_{\nu}$ is the length of the edge in the direction~$\nu$. The continuity of $\phi$ ensures $\phi(\mathbf{r}_0^-) = \phi(\mathbf{r}_0^+) = \phi(\mathbf{r}_0)$. Moreover, $\varepsilon(\mathbf{r}_0^-) = \varepsilon_1$ and $\varepsilon(\mathbf{r}_0^+) = \varepsilon_2$. 
\\

If we now subtract term by term Eq. \eqref{eq:taylor2} from \eqref{eq:taylor1} while keeping only the linear terms and divide both sides by $h_\nu$, we obtain the expression for the incremental ratio of the potential with respect to the grid spacing, which in normal conditions of uniform medium, is the finite difference approximation of the electric field:

\begin{align}
\label{eq:effective_E}
     & -\dfrac{\phi(\mathbf{r}_2) -\phi(\mathbf{r}_1)}{h_{\nu}}= E_{t\nu}(\mathbf{r}_0)+\dfrac{D_{n\nu}(\mathbf{r}_0)}{\varepsilon^{eff}} \mathrel{=:}  E^{eff}_{\nu},
\end{align}
where $\varepsilon^{eff}$ takes the form of the weighted harmonic average (WHA) of $\varepsilon_1$ and $\varepsilon_2$:
\begin{equation}
    \varepsilon^{eff} = \dfrac{1}{\dfrac{\alpha}{\varepsilon_1} + \dfrac{(1-\alpha)}{\varepsilon_2}} \; .
    \label{eq:WHA}
\end{equation}
While harmonic averages have already been used in FEM at the crossing of a discontinuity \cite{robey_pm,ROBERTS1991523} this derivation provides a robust physical interpretation, going beyond the intuitive representation of a series of two capacitors, and a more accurate expression in terms of a WHA.

Thus, $\mathbf{E}^{eff}$ is the electric field corresponding to the correct voltage drop between $\mathbf{r}_1$ and $\mathbf{r}_2$, its tangential component equals that of the actual electric field at $\mathbf{r}_0$.
The corresponding effective electric displacement is $\mathbf{D}^{eff} \mathrel{:=} \varepsilon^{eff}\mathbf{E}^{eff}=\mathbf{D_n}(\mathbf{r}_0)+\varepsilon^{eff}\mathbf{E_t}(\mathbf{r}_0)$, its normal component equals that of the actual electric displacement at $\mathbf{r}_0$.
This construction provides a convenient expression for the dielectric constant to be used in the solution scheme, with an expression which is only slightly more complex than the conventional one and tends to it when $\alpha$ tends to $0$ or to~$1$.

Moreover, if we assume to know $\phi(\mathbf{r}_1)$ and $\phi(\mathbf{r}_2)$, we can derive an interesting expression for the potential located exactly at the intersection with the surface:
\begin{equation}
\begin{aligned}
    \phi(\mathbf{r}_0) & = (1-\alpha)\phi(\mathbf{r}_1) + \alpha\phi(\mathbf{r}_2)+ (1-\alpha)\alpha h_{\nu} D_{n\nu} \Bigl( \dfrac{1}{\varepsilon_2} -\dfrac{1}{\varepsilon_1} \Bigr)  \\
    &\approx \phi(\mathbf{r}_1) + \dfrac{\alpha}{\varepsilon_1} \cdot\dfrac{\phi(\mathbf{r}_2) - \phi(\mathbf{r}_1)}{\dfrac{\alpha }{\varepsilon_1} + \dfrac{(1-\alpha)}{\varepsilon_2}}  \;. \label{eq:phi0}
\end{aligned}
\end{equation}
Expression in Eq.~\eqref{eq:phi0} descends from the following quite reasonable approximation: $D_{n\nu} \approx D^{eff}_{n\nu}$.
Details on its theoretical accuracy are deferred to future work. Here, we will benchmark this choice by comparing the surface potential and the ionic energy, which depends on it, against analytical values on single and multiple-sphere systems.

\subsection{Electrostatic energy calculation}
It is convenient now to recast the energy of a linearized Poisson-Boltzmann system as a sum of three terms, as shown in Eq. \eqref{eq:energyPart}. The Coulombic term, deprived of the singularity related to the self-energy, can be calculated analytically as follows:
\begin{equation}
    E_{coul} = \sum_{i=1}^{N_{atoms}}\sum_{j<i} \dfrac{q_iq_j}{4\pi\varepsilon_m r_{ij}}
\end{equation}
where $r_{ij}$ represents the distance between charges $i$ and $j$. \\
The two remaining energy terms involve surface integrals, and we address them differently.

\subsubsection{The polarization contribution}
The polarization contribution to the electrostatic energy can be written as the energy of the fixed point charges $i$ inside the solute subjected to the potential generated by the polarization charges spread over the MS, presented in Eq.~\eqref{eq:phi_pol}:
\begin{equation}
    E_{pol} = \frac{1}{2}\sum_{i=1}^{N_{atoms}}q_i
    \int_{\Gamma} \dfrac{\sigma_{pol} (\tilde{\mathbf{r}})}{4\pi \varepsilon_0 \| \tilde{\mathbf{r}}-\mathbf{r}_i \|}  \mathrm{d}S
    \label{eq:polen}
\end{equation}
where $\mathbf{r}_i$ is the position of the charge on the $i-$th atom and $\sigma_{pol}$ is the surface polarization charge.

\begin{figure}[h!]
    \centering
    \begin{tikzpicture}
\begin{axis}[
  view={39}{15},
  axis lines=center,
  width=15cm,height=15cm,  xmin=-10,xmax=10,ymin=-10,ymax=10,zmin=-10,zmax=10,
  ticks=none,
  y axis line style=-,
  x axis line style=-,
  z axis line style=-,
]

% plot face flux

\addplot3[
surf,
domain=-5:5,
domain y=-5:5,
color=red, 
opacity=0.01,
fill opacity=0.9, 
faceted color=red!30
](x,{max(y,-x)},{3.5});

%surface
\addplot3[
surf,
domain=-5:5,
domain y=-5:5,
samples=40,
samples y=40,
color=cyan, 
opacity=0.01,
fill opacity=0.9, 
faceted color=cyan!60
]{sin(15*(y+x))+3.5};
\addplot3[
surf,
domain=-5:5,
domain y=-5:5,
color=red, 
opacity=0.01,
fill opacity=0.9, 
faceted color=red!30
](x,{min(y,-x)},{3.5});

\addplot3 [only marks,green,mark=x, very thick] coordinates {(0,0,3.5)};
% Plot one of the lines of intersection
\addplot3 [thick, color=green, dashed] coordinates {
  (-5, 5, 3.5)
  (5, -5, 3.5)
};
\addplot3 [ color=black] coordinates {
  (0, 0, 0)
  (0, 0, 10)
};
% plot dots for the two points

\addplot3 [only marks] coordinates {(0,0,0) (-10,0,0) (10,0,0) (0,-10,0) (0,10,0) (0,0,-10) (0,0,10)};

% plot dashed box
\addplot3 [no marks,densely dashed] coordinates {(5,-5,-5) (-5,-5,-5) (-5,5,-5) (5,5,-5) (5,-5,-5)}; %bottom
\addplot3 [no marks,densely dashed] coordinates {(5,-5,5) (-5,-5,5) (-5,5,5) (5,5,5) (5,-5,5)}; %top
\addplot3 [no marks,densely dashed] coordinates {(5,-5,-5) (5,-5,5) };
\addplot3 [no marks,densely dashed] coordinates {(-5,5,-5) (-5,5,5) };
\addplot3 [no marks,densely dashed] coordinates {(-5,-5,-5) (-5,-5,5) };
\addplot3 [no marks,densely dashed] coordinates {(5,5,-5) (5,5,5) };

% label points
\node [below left] at (axis cs:0,0,0) {$0$};
\node [below] at (axis cs:-10,0,0) {$2$};
\node [below] at (axis cs:10,0,0) {$1$};
\node [below] at (axis cs:0,-10,0) {$3$};
\node [below] at (axis cs:0,10,0) {$4$};
\node [below] at (axis cs:0,0,-10) {$5$};
\node [above] at (axis cs:0,0,10) {$6$};

\end{axis}
\end{tikzpicture}
    \caption{Black dots represent the "current" node $0$ and its six adjacent ones. The green cross represents the intersection between the edge connecting nodes $0$ and $6$ and the MS (in light blue). The reddish square is the intersection between the cube and the plane orthogonal to the edge and passing through the said intersection. The green dashed line represents the intersection between this plane and the MS.}
    \label{fig:intersection}
\end{figure}
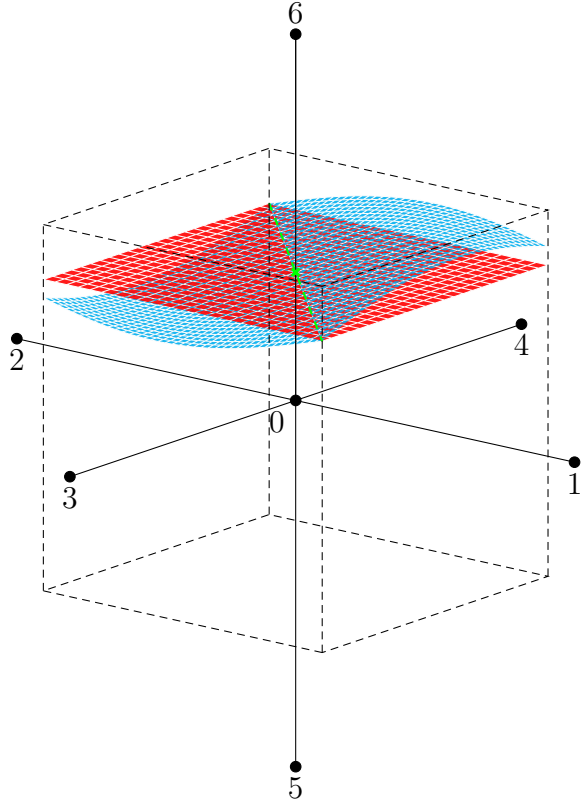

We take advantage of the Gauss law's property to calculate this contribution. In Eq. \eqref{eq:effective_E}, we conjecture that the best way to approximate a region where there is a dielectric discontinuity with a region where the dielectric is uniform is to assume that the uniform dielectric takes the WHA form. We therefore use $\mathbf D^{eff}$ as the best way to estimate the flux of $\mathbf{D_n}$ passing through the MS intersected by a cube. Once we have solved the PBE, we can easily derive, for each MS-intersecting edge, the quantity $D^{eff}_{\nu}h_{\nu}^2$, which corresponds to the flux of $\mathbf D^{eff}$ through a square orthogonal to the edge. 

If we now consider the flux of $\mathbf{D_n}$ through the part of the MS contained in the cube, one can note that an accurate evaluation would require a good knowledge of both $D_n$ and of the local MS shape. But if we apply Gauss law to the space region inside the cube, which is also located between the MS and the square orthogonal to the edge which passes through the intersection, as represented in Fig.~\ref{fig:intersection}, we can suggest, since in our model there is no free charge located around the MS, the following approximation:
\begin{equation}
    D_{n \nu}h_{\nu}^2 \big |_{square} \approx \iint_{MS \cap cube} D_{n} dS \: .
    \label{Dnvh_approx_}
\end{equation}

% \begin{figure}
%     \centering
%     \begin{tikzpicture}
% \filldraw[color=blue!60, ultra  thick] node[my_rectangle,canvas is zy plane at x=2, fill=blue!5] {};
% \draw [black, thick] (-0.5,0) --(4.5,0);
% \draw [black, thick] (0,-0.5) --(0,0.5);
% \draw [black, thick] (4,-0.5) --(4,0.5);
% \filldraw [black, ultra thick] (0,0) circle (2pt);
% \filldraw [black, ultra thick] (4,0) circle (2pt);
% \draw [red] (0.8,0) node {X};
% \draw [blue] (2,1) node {$\Gamma_j$};
% \filldraw [color=blue!60, thick] (2,0) circle (2pt);
% \end{tikzpicture}
%     \caption{Surface of the cube grid face perpendicular to an intersected edge. The red cross represents the position of the intersection and the blue dot represents the edge midpoint.}
%     \label{fig:intersection}
% \end{figure}

By means of this approach, we derive the total polarization charge $q_p \approx \varepsilon_0 \left(\dfrac{1}{\varepsilon_s} - \dfrac{1}{\varepsilon_m}\right) D^{eff}_{\nu}h_{\nu}^2$  located on the piece of MS intersecting a cube and concentrate it on the intersection point $\mathbf{r}_p$ between the MS and the grid. 
The final calculation takes, therefore, the following form:
\begin{equation}
 E_{pol} = \frac{1}{2}\sum_{i=1}^{N_{atoms}}
    \sum_{p=1}^{N_{Ic}} \dfrac{q_i q_p}{4\pi \| {\mathbf{r}_p}-\mathbf{r}_i \|}
\end{equation}
where $N_{Ic}$ is the total number of cubes intersecting the MS.

\subsubsection{Ionic direct contribution calculation}
\label{sect:ionicCalc}
\begin{figure}
    \centering
    \includegraphics[width=1\linewidth]{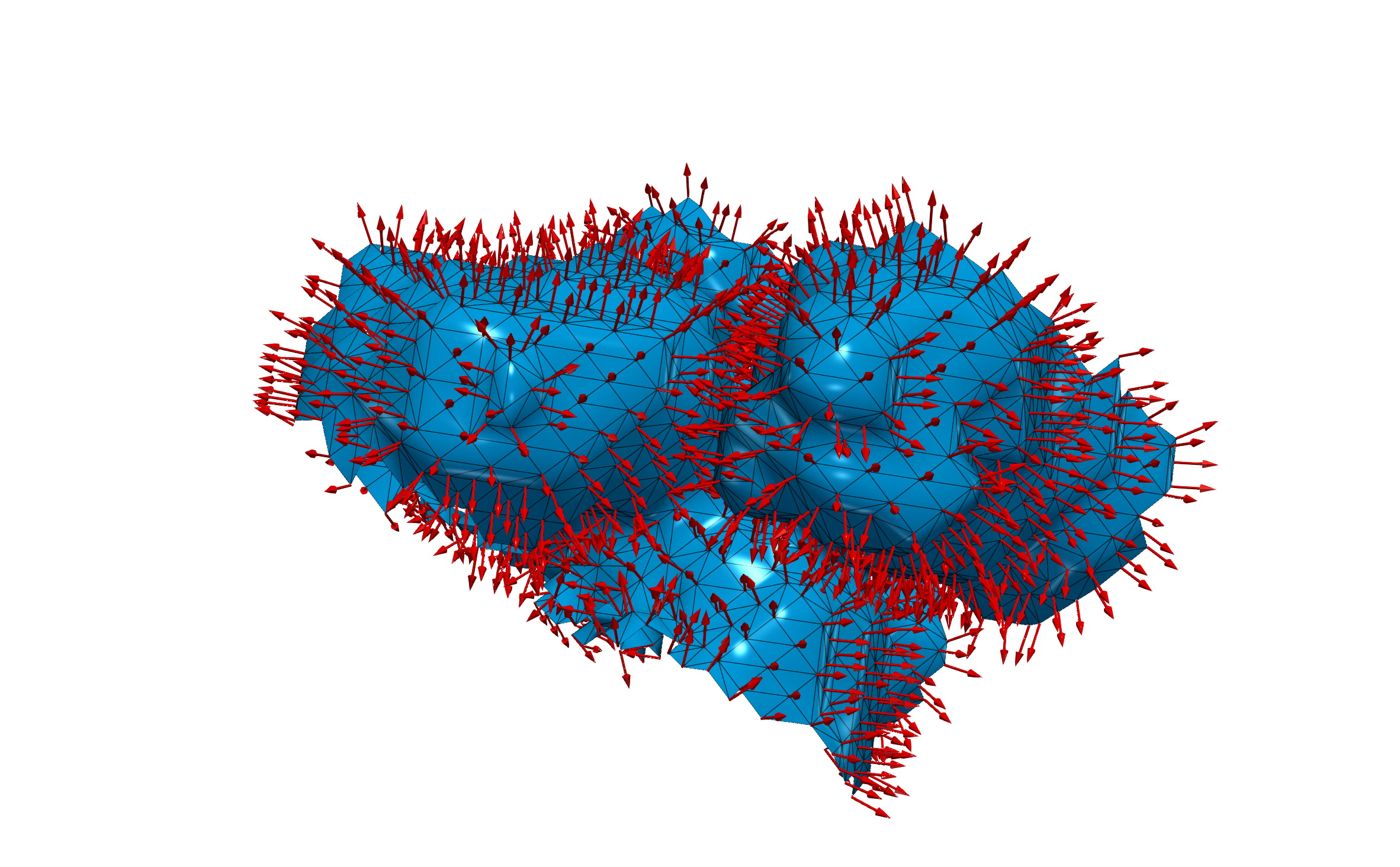}

    \caption{Triangulation of the MS with normal vectors given by NS.}
    \label{fig:triangulation}
\end{figure}

By considering Eqs. \eqref{eq:pot_inside_final} and \eqref{eq:energyPart}, one can get the expression for the direct ionic contribution:
\begin{equation}
    E_{ion} = \frac{1}{2}\sum_{i=1}^{N_{atoms}}q_i\left(\int_{\Gamma} \phi(\tilde{\mathbf{r}}) 
    \dfrac{(\tilde{\mathbf{r}}-\mathbf{r}_i) \cdot \mathbf{n}(\tilde{\mathbf{r}})}{4\pi \| \tilde{\mathbf{r}}-\mathbf{r}_i \|^3}  \mathrm{d}S 
    - \dfrac{1}{\varepsilon_s}\int_{\Gamma} \dfrac{\mathbf{D}(\tilde{\mathbf{r}}) \cdot \mathbf{n}(\tilde{\mathbf{r}})}{4\pi \| \tilde{\mathbf{r}}-\mathbf{r}_i \|}  \mathrm{d}S \right)
    \label{eq:energy_ion}
\end{equation}
Here,  the second integral is equivalent, up to a multiplicative constant, to that in Eq. \eqref{eq:polen}, and thus is treated similarly.\\
We now focus on discretizing the first integral of Eq. \eqref{eq:energy_ion}:

\begin{equation}
    I = \int_{\Gamma} \phi(\tilde{\mathbf{r}}) 
    \dfrac{(\tilde{\mathbf{r}}-\mathbf{r}_i) \cdot \mathbf{n}(\tilde{\mathbf{r}})}{4\pi \| \tilde{\mathbf{r}}-\mathbf{r}_i \|^3}  \mathrm{d}S \;.
    \label{eq:int_2}
\end{equation}
To evaluate this integral, we take again the advantage of the information delivered by NanoShaper, namely analytical intersections and normals (Fig.~\ref{fig:triangulation}). This is used to locally re-triangulate the surface via the marching cubes algorithm. The total molecular surface is hence approximated as the union of the triangles:
\begin{equation}
    \Gamma \approx \bigcup_{j=1}^{N_{t}} \Gamma_j 
\end{equation}
where $N_{t}$ is the number of triangles in the cube grid.
Overall, using this information, the integral \eqref{eq:int_2} can be discretized as follows:

\begin{equation} 
I \approx \sum_{j=1}^{N_{t}} \int_{\Gamma_{j}} \phi(\tilde{\mathbf{r}}) \dfrac{(\tilde{\mathbf{r}}-\mathbf{r}_i) \cdot \mathbf{n}(\tilde{\mathbf{r}})}{4\pi \| \tilde{\mathbf{r}}-\mathbf{r}_i \|^3} \mathrm{d}S 
\approx 
\sum_{j=1}^{N_{t}}\dfrac{T_j}{3}\sum_{k=A,B,C}\phi(\mathbf{r}_k) \dfrac{(\mathbf{r}_k-\mathbf{r}_i) \cdot \mathbf{n}(\mathbf{r}_k)}{4\pi \| \mathbf{r}_k-\mathbf{r}_i \|^3}
\end{equation}
where $T_j$ is the area of the $j$-th triangle and $k$ labels its three vertices $A$, $B$ and $C$. The potentials at the triangle vertices are calculated through the formula~\eqref{eq:phi0}. 

While the type of calculations performed in Sect.~\ref{sect.Partitioning} are quite common in the treatment of elliptic PDEs, we are not aware of any PB solver implementing this solution for the calculation of the ionic contribution to the electrostatic energy.

\subsection{Exploiting de-refinement for more efficient BCs}
\label{sect:New_BCs}

We have shown in Sect.~\ref{sec:BCs} that even the likely most accurate BCs used in Finite Difference and Finite Element methods, often called Coulombic, cannot be considered asymptotic, and their accuracy may be sensitive to particular characteristics of the computed system. Furthermore, their computational cost can become significant for large systems, as it involves calculating distance-based functions between all atomic charges and all elements at the boundary of the computational domain where the equation is solved. On the other side, null, homogeneous Dirichlet BCs would, in principle, be exact at a very far distance from the solute regardless of its features, but using a very low percentage filling (aka perfil) could be computationally unfavorable, wasting a lot of computational effort on degrees of freedom located far in the solvent.
Let's consider the possibilities given by numerical advances, though. We may note that grid adaptivity, which is usually adopted to increase the accuracy in the most critical regions, which in the PB are the point charges and the MS, can very simply be applied to reduce the resolution in the regions where there are no criticalities. Moreover, while changing resolution near complex geometries may have inherent complexities beyond the simple number of d.o.f., doing this in the solvent is much lighter. This is why we suggest that the overall best trade-off between accuracy and computational cost is to use a uniform grid tight around the system, say with a perfil around $90\%$, followed by a de-refinement or the grid, until a much lower perfil is reached, say between $15\%$ and $20\%$. It is worth noting that, due to the specific octree structure of our grid, if we fix the "tight" box size, we can no longer exactly determine the perfil within the entire computational domain. Indeed, the octree structure of cubic elements imposes that the entire computational domain is again a cube having a side length, which is a power of 2 of the inner grid spacing. So, one can only fix an upper bound for the overall perfil.

\section{Software Architecture}
\label{sect:solver_structure}
The philosophy that guided the design of NextGenPB was, on the one hand, to rely as much as possible on existing Free Software libraries that provide the required capabilities, whenever possible, while, on the other hand, integrating such libraries by means of \emph{adapters} providing an intermediate level of abstraction allowing, when needed, the plug-in addition or  replacement of  libraries serving similar purposes. This approach is indeed inherited in NextGenPB from its main dependency which is the bimpp C++ library~\cite{bimpp,bimpp_website}. 
bimpp provides the main methods that implement the distributed memory parallel assembly
of the discrete operators discussed in Sect.~\ref{sect:primal_mixed} and, in addition, adapters for
parallel solvers for the related linear systems via direct~\cite{amnestoy_mumps,mumps_website} or iterative~\cite{nishida_lis,lis_website} methods. 
While many possible solvers can be invoked through the available interfaces, given the structure
of the system derived from the discretization of the linearized PB equation, the most used solver
in our tests is the Conjugate Gradient solver implemented in LIS with a SSOR preconditioner so,
unless stated otherwise, one can assume that this is the choice adopted for all the tests in the following sections.
bimpp also provides a class for managing hierarchically (de)refinable 
Cartesian Oct-tree meshes, which is essentially a wrapper 
around the C library p4est~\cite{burstedde_p4est,p4est_website}.\footnote{While the initial plan was to implement this wrapper as a generic interface allowing for the replacement of p4est with other Octree libraries, the excellent performance of p4est has made the implementation of additional adapters a very low priority task.}
The description of the molecular surface is handled through an interface to the NanoShaper library which implements a set of different surface description formats~\cite{NanoShaper,NanoShaper-vmd}. 
Parallel I/O of structured binary data is implemented in bimpp via linking to the liboctave library which is part of GNU Octave~\cite{octave}, so that the output files can be conveniently post processed via scripts using the GNU Octave interpreter; output in vtk file format is also available for visualization.

\section{An analytical benchmark}
\label{analytical_benchmark_section_label}
For the case of solutes represented by $N_s$ non-overlapping dielectric spheres $\Omega_{m,i}$ with the same relative dielectric constant $\varepsilon_{r,m}$ and centered at points $\mathbf r_i\in\mathbb R^3$ (each sphere $\Omega_{m,i}$ is characterized by a radius $R_i$ and contains a fixed centrally-located point charge $q_i$, $i=\overline{1,\ldots,N_s}$) we look for the total self-consistent potential $\phi(\mathbf r)$ (at a point $\mathbf r\in\mathbb R^3$) in the form~\cite{our_jcp, our_jpcb}
\begin{equation}
\label{potential_total_selfconsist_}
\phi(\mathbf r) = \begin{cases}
\phi_{\text{in},i}(\mathbf r) = \phi_{coul,i}(\mathbf r) + \Check\phi_{\text{in},i}(\mathbf r), &  \mathbf r\in\Omega_{m,i},\\
\phi_{\text{out}}(\mathbf r) = \sum_{i=1}^{N_s} \phi_{\text{out},i}(\mathbf r), & \mathbf r\in \Omega_s,
\end{cases}
\end{equation}
where the sum $\sum_{i=1}^{N_s} (\cdot)$ in $\phi_\text{out}$ reflects the superposition principle applicable to the Debye-H\"uckel description. Addend $\phi_{coul,i}(\mathbf r) = \frac{q_i}{4\pi\varepsilon_0\varepsilon_{r,m} \|\mathbf{r} - \mathbf r_i\|}$ in \eqref{potential_total_selfconsist_} is obviously the Coulombic potential due to the given free charge $q_i$ situated at $\mathbf r_i$, while unknown potentials $\Check\phi_{\text{in},i}$ and $\phi_{\text{out},i}$ are expressed through local eigenfunction expansions (of Laplace type for $\Check\phi_{\text{in},i}$ and Poisson-Boltzmann type for $\phi_{\text{out},i}$)
\begin{equation}
\label{Lin_pbe_phi_in_out}
\begin{aligned}
\Check\phi_{\text{in},i}(\mathbf r) & = \sum\nolimits_{n,l}L_{n l,i} \Tilde\varrho_i^n Y_n^l(\Hat{\boldsymbol\varrho}_i), \\
\phi_{\text{out},i}(\mathbf r) & = \sum\nolimits_{n,l}G_{n l,i}k_n(\Tilde\varrho_i) Y_n^l(\Hat{\boldsymbol\varrho}_i),
\end{aligned}
\end{equation}
where dimensionless $\Tilde\varrho_i = k_D \varrho_i$ is the scaled (by a Debye screening length $k_D^{-1}$) radial coordinate of $\mathbf r$ measured from the $i$-th sphere's center (i.e.~$\boldsymbol\varrho_i = \mathbf r-\mathbf r_i$, $\varrho_i = \|\boldsymbol\varrho_i\|$, $\Hat{\boldsymbol\varrho}_i = \boldsymbol\varrho_i/\varrho_i$), $k_n(\cdot)$ are modified spherical Bessel functions of the 2nd kind, $Y_n^l(\cdot)$ are the standard orthonormal complex-valued spherical harmonics (see \cite{Jack}), and $\sum_{n,l}(\cdot)$ denotes the sum $\sum_{n=0}^{+\infty}\sum_{l=-n}^n(\cdot)$. Unknown coefficients $L_{n l,i}$ and $G_{n l,i}$ of \eqref{Lin_pbe_phi_in_out} are to be determined from boundary conditions on the spheres' boundaries (see details in~\ref{analytical_benchmark_appendix_label}).

According to \eqref{potential_total_selfconsist_}, the total electrostatic energy can now be calculated as $E = \frac{1}{2}\sum_{i=1}^{N_s} q_i \cdot \left.(\phi_{\text{in},i}-\phi_{coul,i})\right|_{\varrho_i=0} = \frac{1}{2}\sum_{i=1}^{N_s} q_i\cdot\left.\Check\phi_{\text{in},i}\right|_{\varrho_i=0} = \frac{1}{2}\sum_{i=1}^{N_s} q_i L_{0 0,i}/\sqrt{4\pi}$ (note that $Y_0^0 = \frac{1}{\sqrt{4\pi}}$); subtracting the Coulombic potential $\phi_{coul,i}$ from the full potential $\phi_{\text{in},i}$ (see \eqref{potential_total_selfconsist_}), to avoid infinity in energy, corresponds to the removal of the so-called self-energy term~\cite{rocchia2001extending,our_jcp}. 

To deepen the possibilities of comparing calculations based on this analytical model with the results of numerical calculations \cite{rocchia2001extending}, it is desirable to be able to calculate not only the total electrostatic energy $E$, but also its components (see \eqref{eq:energyPart}). While the calculation of the Coulombic part is trivial, calculations of other parts become more involved -- e.g.~in the considered case of centrally located point charges, one can show (see details in~\ref{analytical_benchmark_appendix_label}) that the corresponding resulting reaction potential $\phi_{\text{pol},i,i}$ at point $\mathbf r_i$ (i.e.~at the $i$-th sphere's center) created by the total polarization charge density on the $i$-th surface ($\varrho_i=R_i$)~is
\begin{equation}
\label{phi_pol_ii}
\phi_{\text{pol},i,i} = \frac{q_i}{4\pi\varepsilon_0 R_i}\!\left( \frac{1}{\varepsilon_{r,s}} - \frac{1}{\varepsilon_{r,m}}\right) \! ,
\end{equation} 
while the potential $\phi_{\text{pol},i,j}$ created by the same density at point $\mathbf r_j$ (i.e.~at the $j$-th sphere's center with $j\ne i$)~is
\begin{equation}
\label{phi_pol_ij}
\begin{aligned}
\phi_{\text{pol},i,j} = & \frac{q_i}{4\pi\varepsilon_0 a_{i j}}\!\left( \frac{1}{\varepsilon_{r,s}} - \frac{1}{\varepsilon_{r,m}}\right) \\ 
& + \left(1 - \frac{\varepsilon_{r,m}}{\varepsilon_{r,s}}\right)\sum_{n,l} \frac{n\Tilde R_i^n L_{n l,i}}{2 n+1} \left(\frac{R_i}{a_{i j}}\right)^{\! n+1} Y_n^l(\Hat{\mathbf a}_{i j}),
\end{aligned}
\end{equation} 
$\Tilde R_i = k_D R_i$, $\mathbf a_{i j}=\mathbf r_j-\mathbf r_i$ (see \ref{analytical_benchmark_appendix_label} for details). With the reaction potentials \eqref{phi_pol_ii} and \eqref{phi_pol_ij} so obtained one can immediately calculate the corresponding (polarizational) energy contributions $\frac{1}{2} q_i \phi_{\text{pol},i,i}$ and $\frac{1}{2} q_j \phi_{\text{pol},i,j}$, for all $i, j = \overline{1,\ldots,N_s}$, $j\ne i$; note that $\frac{1}{2} q_i \phi_{\text{pol},i,i} = \frac{q_i^2}{8\pi\varepsilon_0 R_i}\bigl( \frac{1}{\varepsilon_{r,s}} - \frac{1}{\varepsilon_{r,m}}\bigr)$, which coincides with the conventional Born energy of a single sphere at zero ionic strength (see~\cite{our_jcp}). Ionic energy contributions (see~\eqref{eq:energyPart}), as they appear in \cite{rocchia2001extending}, can then be calculated by subtracting the Coulombic part and the (just derived) polarization contributions from the total energy~$E$. \\
We have written MATLAB scripts to calculate the potential and energy using the above analytical expressions (\url{https://github.com/concept-lab/Analytical_Electrostatics/tree/main}).

%% file: Results.tex
\section{Results}
\label{sec:results}

This section provides a detailed accuracy and computational cost analysis, comparing the proposed solver with the leading solvers commonly used in biomolecular simulations. The benchmarks include both analytical solutions and real biomolecular systems. Additionally, we present results on parallelization and scalability obtained using High-Performance Computing (HPC) architectures.\\
Unless otherwise specified, the following simulations use a solvent dielectric constant of 80, a solute dielectric constant of 2, an ionic strength of 0.145~$mol/L$ and a temperature of 298.15K.

\subsection{Accuracy}

\subsubsection{Analytically Solvable Systems}
The comparison is made against the finite difference (FD) calculation implemented in the DelPhi \cite{DELPHI} solver and the second-order accuracy implemented in the MIBPB solver \cite{MIBPB, MIBPB2}. The ground truth consists of the analytical calculations for a many-sphere system, as described in Sect. \ref{analytical_benchmark_section_label} and having foundations in \cite{Yu3,Yu2021,our_jcp,our_jpcb,Yu2019} and references therein. To perform a fair comparison, the same grid structure and parameters are used in the benchmarks despite the greater flexibility of the NGPB solver.

\begin{figure}[!h]
    \centering
    \includegraphics[width=0.8\linewidth]{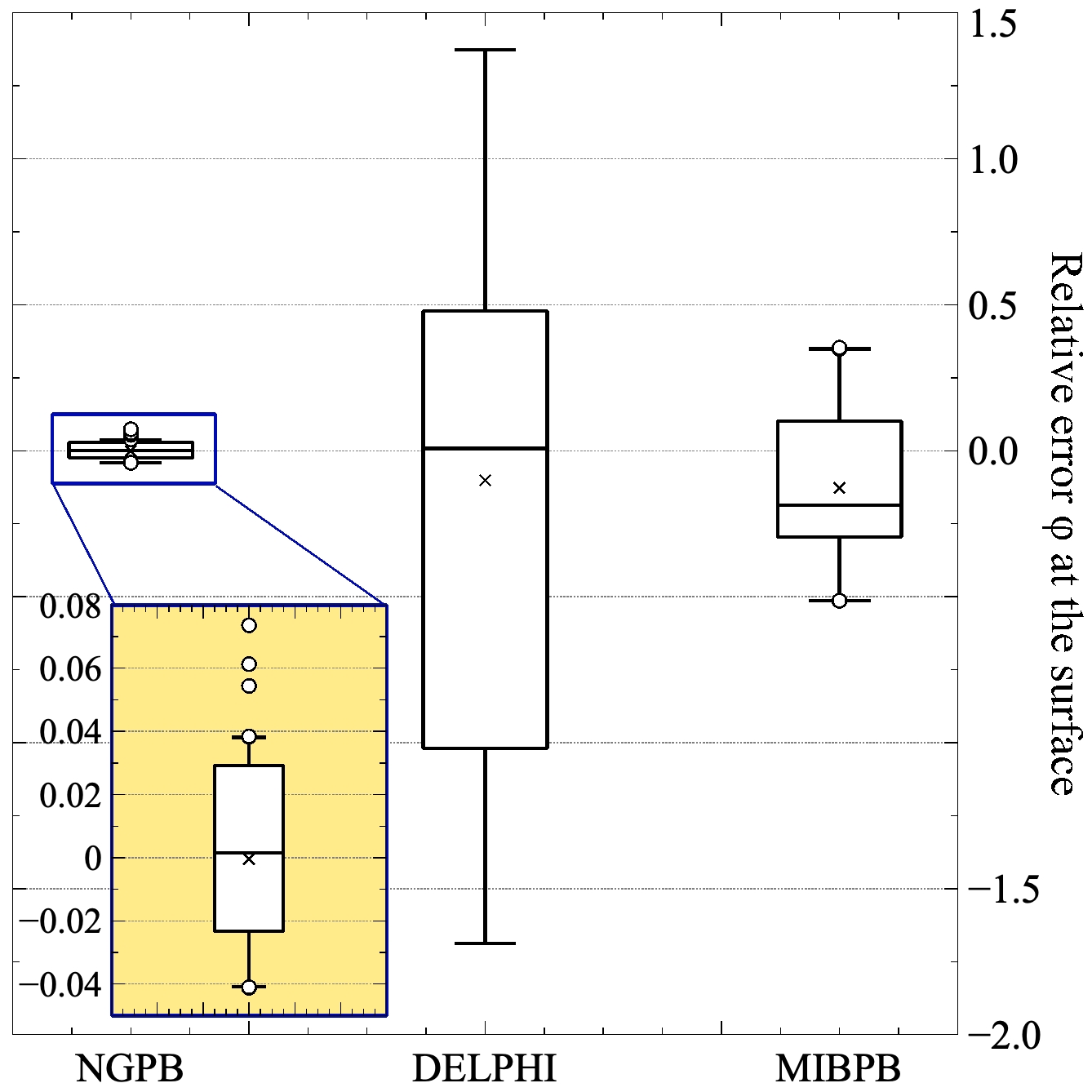}
    \caption{Box plot of relative error for the electrostatic potential at the surface compared against analytical results for NGPB, DelPhi and MIBPB for the Kirkwood sphere. }
    \label{fig:1_sfera_confronto_solver}
\end{figure}

\paragraph{The Kirkwood sphere}
First, we consider a simple system for which the linearized Poisson-Boltzmann equation (LPBE) can be analytically integrated: a single dielectric sphere immersed in an electrolytic solution. Although the spherical system can easily be solved analytically, it shares some commonalities with more complex alternatives since its symmetry differs from that of the Cartesian grid on which it is mapped and solved. The analytical expressions for the polarization and ion energies are~\cite[Eq.~(88)]{our_jcp}:
%\begin{equation}
\begin{align}
    E_{pol} &= \dfrac{1}{2}\left(\dfrac{1}{\varepsilon_s} -\dfrac{1}{\varepsilon_m} \right)
    \dfrac{q^2}{4\pi R} \\
    E_{ion} &= -\dfrac{1}{2}\dfrac{q^2}{4\pi\varepsilon_s} \dfrac{k_D}{(1+k_D R)}
\end{align}
%\end{equation}
where $q$ is the charge, $R$ is the radius of the sphere, and $k_D$ is the inverse Debye length. We use a uniform mesh with a grid spacing of 0.5~\AA~for all solvers, with 15\% of the box filled and homogeneous Dirichlet boundary conditions. This setup isolates the effect of boundary conditions from the calculations. The sphere has a radius of 2~\AA~and a charge of 1 e.s.u. We compare the energy and the potential at the surface of each solver with the analytical solution. As shown in Table \ref{tab:energies_1sf}, the relative error for NGPB is at least one order of magnitude lower than that of the other solvers for each energy term. To represent the accuracy of the potential at the surface, interpolating the electrostatic potential at any surface point inside a cubical volume using the potential values at the eight vertices employing a trilinear function, we use a box plot (Fig. \ref{fig:1_sfera_confronto_solver}). In this plot, the cross represents the mean of the data, the box represents the interquartile range (IQR), containing the middle $50\%$ of the data, and the whiskers extend to cover $96\%$ of the data. The most extreme $2\%$ on either side are considered outliers.

% \begin{table}[!h]
%     \centering
%     \begin{tabular*}{0.95\columnwidth}{lrrr}
%     \hline
%     - & Polarization & Ionic & Total    \\
%     \hline
%     Percentage of box filling 12.5\%    \\
%     DelPhi & -2.25E-07 & -5.10E-01 & -2.59E-03 \\
%     NGPB & 7.38E-10 & 3.36E-02 & 1.70E-04 \\
%     NGPB$^*$ & 7.38E-10 & -2.46E-02 & -1.25E-04\\
%     \hline
%     Percentage of box filling 50\%    \\
%     DelPhi & -2.25E-07 & -5.10E-01 & -2.59E-03 \\
%     NGPB & 7.38E-10 & 3.36E-02 & 1.70E-04 \\
%     NGPB$^*$ & 7.38E-10 & -2.46E-02 & -1.25E-04\\
%     \hline
%     \end{tabular*}%
%     \caption{Signed relative error in energy calculation. (in the NGPB$^*$ calculation, the exact value for the area of each individual tile of the surface is used)}
%     \label{tab:energies_1sf}
% \end{table}

%\sd{i solver usano float o double? usano tutti la stessa precisione?}
%\vr{credo tutti double, ma non ho certezza. PATRIZIO?}
%\pa{apbs double, DElPhi  alexov compilation in double, mibpb non saprei perhè non è documentato i sorgenti non li ho}

\begin{table}[!h]
    \centering
    \begin{tabular*}{0.69\columnwidth}{l|ccr}
    \hline
    - & Polarization & Ionic & Total    \\
    \hline
    Value $[K_b T]$   & -68.31 & -0.35 & -68.65 \\
    \hline
    \multicolumn{4}{c}{Relative error}\\
    \hline
    NGPB   & 7.38e-10 & 3.39e-02 & 1.72e-04 \\
    \hline
    DelPhi & 5.89e-05 & -5.98e-01 &-2.97e-03\\
    \hline
    MIBPB   & \multicolumn{2}{c}{7.16e-03} & 7.16e-03\\
    \hline
    
    \end{tabular*}%
    \caption{Analytical value and signed relative error in energy calculation for NGPB, DelPhi and MIBPB for the Kirkwood sphere. The percentage of box filling is $15\%$. For MIBPB, we could not identify the energy contributions from the polarization and ionic terms individually, so we report the error for their combined total along with the total energy error. }
    \label{tab:energies_1sf}
\end{table}

%%%%%%%%%%%%%%%%%%%%
%% sfera usata è di raggio 3.2 AA

% Using adaptive mesh, we can select a specific area of the computational domain where we set the target grid spacing, and we can de-refine outside that box. In this way, we can enlarge the computational domain without enlarging the number of grid points, see Fig ... . Enlarging the domain can lead us to set the Dirichlet boundary condition with more accuracy w.r.t the Coulombic boundary condition in a small region. For one sphere, we compare using Coulombic BCs at a perfil of 80$\%$
% with null BCs using derefinement from 80$\%$ until 20$\%$ and from 95$\%$ until 20$\%$.
% We managed to improve the BCs' quality and save in degrees of freedom as the results in the table... show.
% \begin{table}[h!]
%     \centering
%     \begin{tabular}{c|c|c|c}
%        \makecell {Mean \\ relative error }
%        & D-H BCs 80$\%$
%        & \makecell {Dirichlet BCs 80$\%$ \\  derefined until 20$\%$}
%        & \makecell {Dirichlet BCs 95$\%$ \\ derefined until 20$\%$} \\
%     \hline
%        \makecell {Node\\ Potential} &
%        0,0344 &
%        0,0262 &
%        0,0292 \\
%     \hline
%        \makecell {Surface\\ Potential} &
%        0,0365 &
%        0,0251 &
%        0,0286
%     \end{tabular}
%     \caption{Caption}
%     \label{tab:my_label}
% \end{table}
%%%%%%%%%%%%%%%%%%%%

\begin{figure}[h!]
    \centering
    \includegraphics[width=1\linewidth]{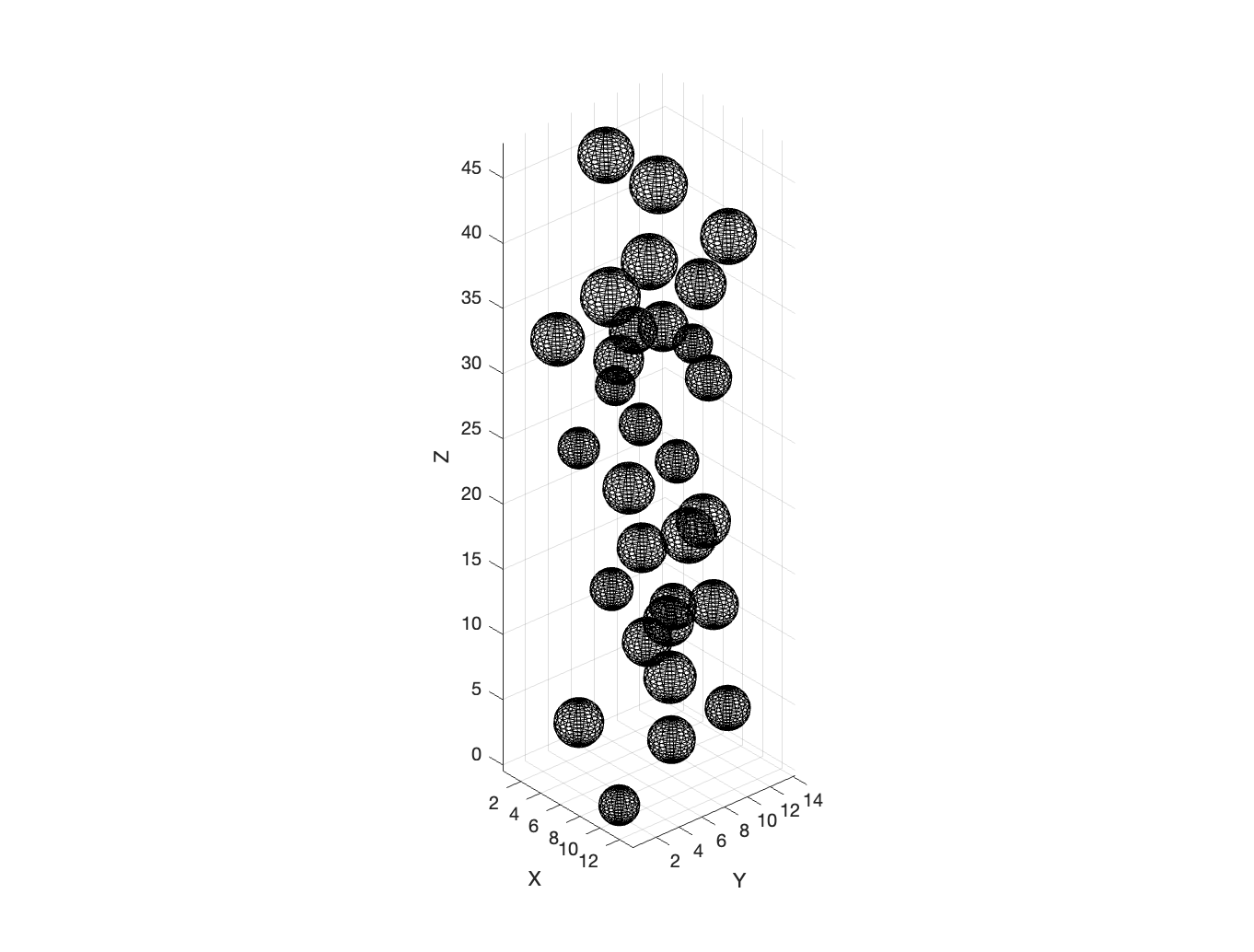}
    \caption{Schematic representation of 30--spheres system.}
    \label{fig:30_sfere}
\end{figure}

\begin{figure}[!h]
    \centering
    \includegraphics[width=0.8\linewidth]{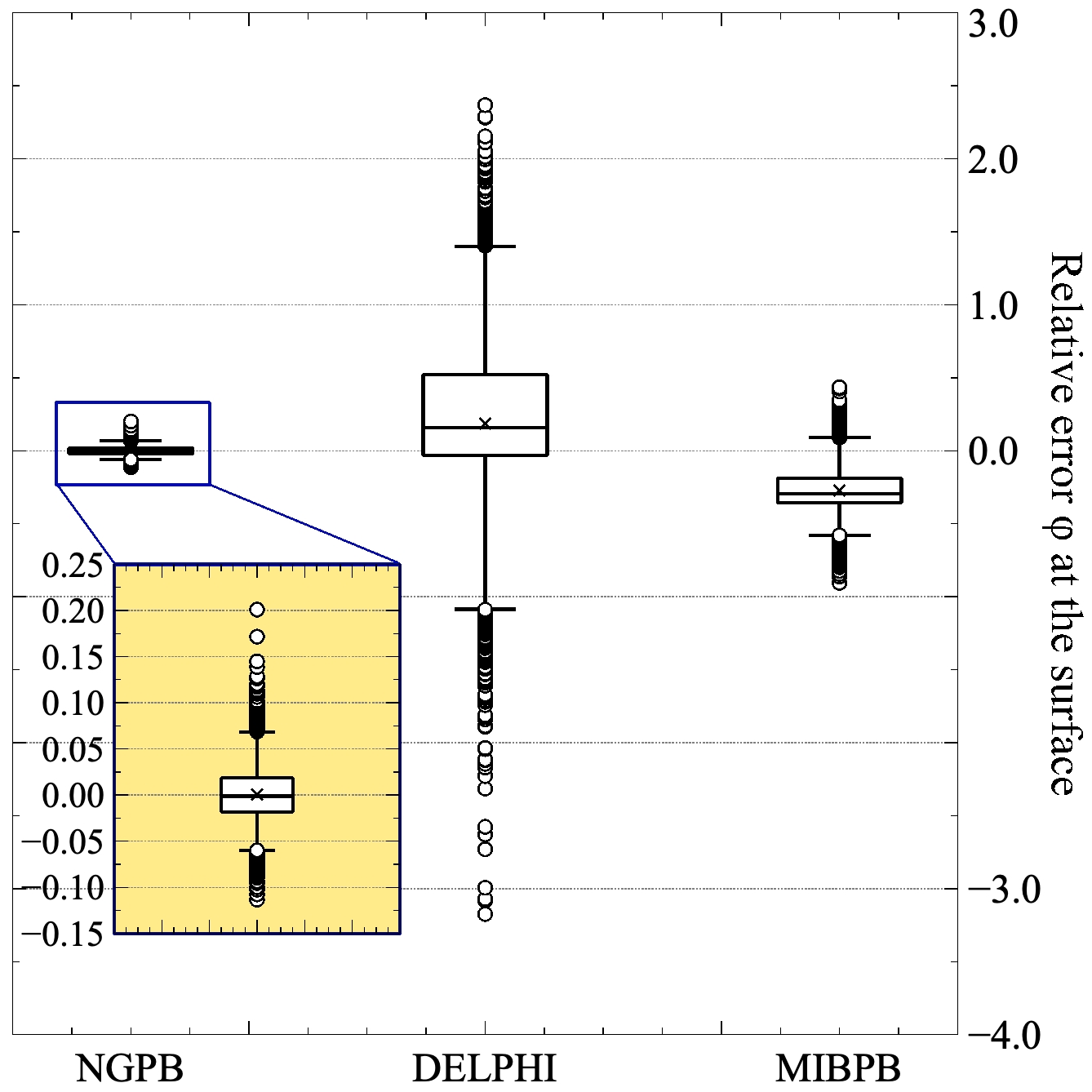}
    \caption{Box plot of relative error of potential at the surface compared with analytical results for NGPB, DelPhi and MIBPB on the 30-sphere system. }
    \label{fig:30_sfere_confronto_solver}
\end{figure}

\paragraph{Non-overlapping Spheres}
We showed in Sect.~\ref{analytical_benchmark_section_label} that it is possible to achieve an analytical solution also for systems composed of many non-overlapping spheres. These systems pose challenges to the PB solution very similar to real biomolecules, except for the construction of the MS. For this purpose, we select a 30-sphere system, as shown in Fig.~\ref{fig:30_sfere}. The positions of the centers of the 30 spheres are chosen to ensure that the system lacks any symmetry. Additionally, the charges of the atoms in the three spheres with the greatest $z$-values are set to zero to evaluate the accuracy of the potential in these atoms, a procedure commonly done in the protocols that estimate the pKa of titratable residues. For this benchmark, we assess our code in terms of accuracy and compare it with that of other well-established solvers, such as DelPhi and MIBPB. For all solvers, we use a uniform mesh with a grid spacing of 0.5~\AA, with 20\% of the box filled and homogeneous Dirichlet boundary conditions.
Results are presented in Tables \ref{tab:energies_30sf}, \ref{tab:pot_30sf} and Fig.~\ref{fig:30_sfere_confronto_solver}. As in the case of a single sphere, NGPB turns out to be remarkably accurate in calculating both energy and potential at the surface and the atom centers.

\begin{table}[!h]
    \centering
    \begin{tabular*}{0.69\columnwidth}{l|ccr}
    \hline
    - & Polarization & Ionic & Total    \\
    \hline
    Value $[K_b T]$   & -10310.57 & -151.13 & -2255.59 \\
    \hline
    \multicolumn{4}{c}{Relative error}\\
    \hline
    - & Polarization & Ionic & Total    \\
    \hline
    NGPB   & -4.16e-05 & 1.39e-02 & 7.46e-04 \\
    \hline
    DelPhi & 4.13e-04 & -8.85e-02 &-4.04e-03 \\
    \hline
    MIBPB &  \multicolumn{2}{c}{4.00e-03} & 1.86e-02\\
    \hline
    
    \end{tabular*}%
    \caption{Analytical value and signed relative error in energy calculation for NGPB, DelPhi and MIBPB on the 30-sphere system. The percentage of box filling is $20\%$, and null BCs are used. For MIBPB, we can not separate the energy contributions from the polarization and ionic terms individually, so we report the error for their combined total along with the total energy error. }
    \label{tab:energies_30sf}
\end{table}

\begin{table}[!h]
    \centering
    \begin{tabular*}{0.6\columnwidth}{lrrr}
    \hline
    - & Atom1 & Atom2 & Atom3    \\
    \hline
    NGPB   & 2.71e-05 & -1.19e-03 & -4.66e-03 \\
    \hline
    DelPhi & 6.04e-04 & 4.60e-03 &8.56e-03 \\
    \hline
    MIBPB   & -2.99e-01 & -3.07e-01 & -2.97e-01\\
    \hline
    
    \end{tabular*}%
    \caption{Signed relative error on uncharged atoms for NGPB, DelPhi and MIBPB on the 30-sphere system. The percentage of box filling is $20\%$ and null BCs are used.}
    \label{tab:pot_30sf}
\end{table}

\begin{figure}[!b]
    \centering
    \includegraphics[width=0.5\linewidth]{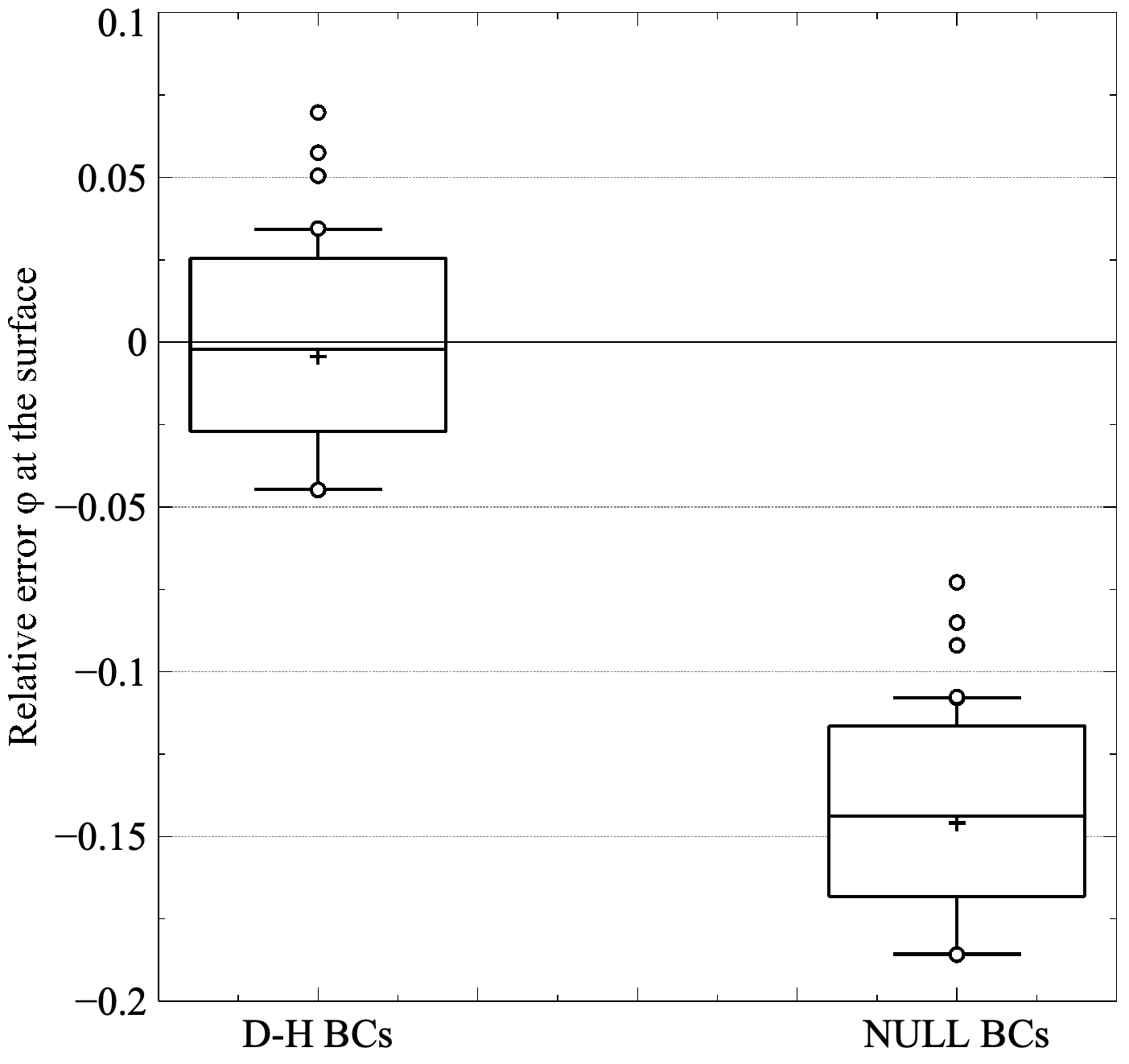}
    \caption{Box plot of relative error of potential at the surface compared with analytical results with NGPB assigning different BCs for one-sphere systems at 80$\%$.}
    \label{fig:conf_1_sfera_BC}
\end{figure}

\paragraph{BCs assessment}
We test the proposed approach for the BCs here, which is described in Sect.~\ref{sect:New_BCs}. 
We first show that, as already observed, the BCs and some grid artifacts impact different outputs of the PB solution's outputs differently. Thus, we compare the differences in energies and potentials on a single-sphere system upon application of D-H/Coulombic BCs wrt null BCs at a perfil of $80\%$. At this perfil value, the former BCs are far more accurate. However, as it can be seen in Fig.~\ref{fig:conf_1_sfera_BC} and in Table~\ref{tab:energy_1_sfera_bc}, this difference in accuracy impacts substantially the potential at the surface and, as a consequence, on the ionic energy term, but it goes unnoticed if only the polarization term, which is the most significant, is considered.

\begin{table}[!b]
    \centering
    \begin{tabular*}{0.7\columnwidth}{lrrr}
    \hline
    BCs & Polarization & Ionic & Total \\
    \hline
    D-H @80\%  & 7.38e-10 & 4.91e-02 & 2.49e-04 \\
    \hline
    NULL @80\% & 7.38e-10 & 6.12e-01 & 3.10e-03 \\
    \hline
    
    \end{tabular*}%
    \caption{Signed relative error in energy calculation with NGPB assigning different BCs for one-sphere systems at 80$\%$.}
    \label{tab:energy_1_sfera_bc}
\end{table}

\begin{figure}[h!]
    \centering
    \includegraphics[width=0.5\linewidth]{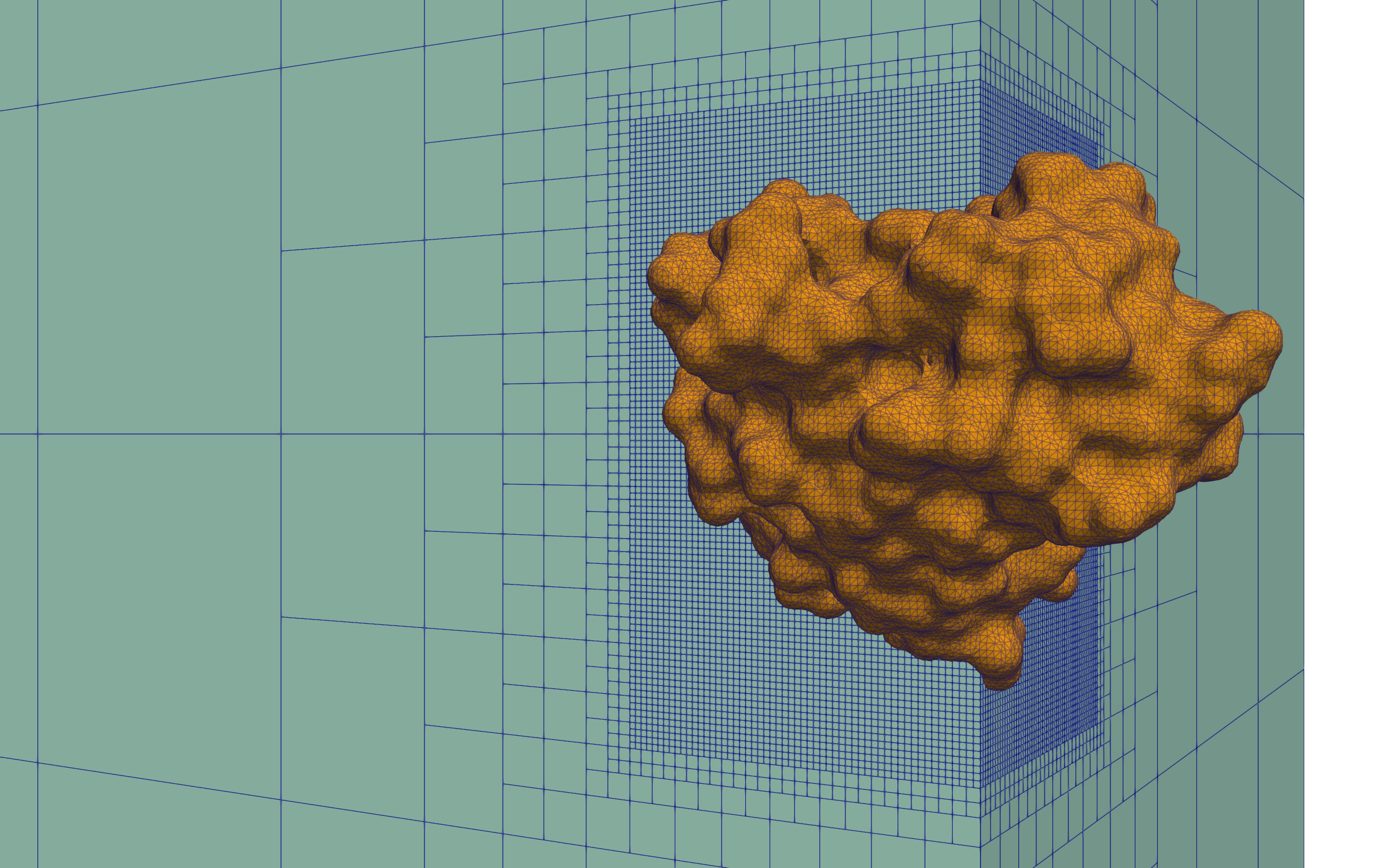}
    \caption{Representation of the adaptive grid.}
    \label{fig:deraffinamento}
\end{figure}

\begin{figure}[h!]
    \centering
    \includegraphics[width=0.5\linewidth]{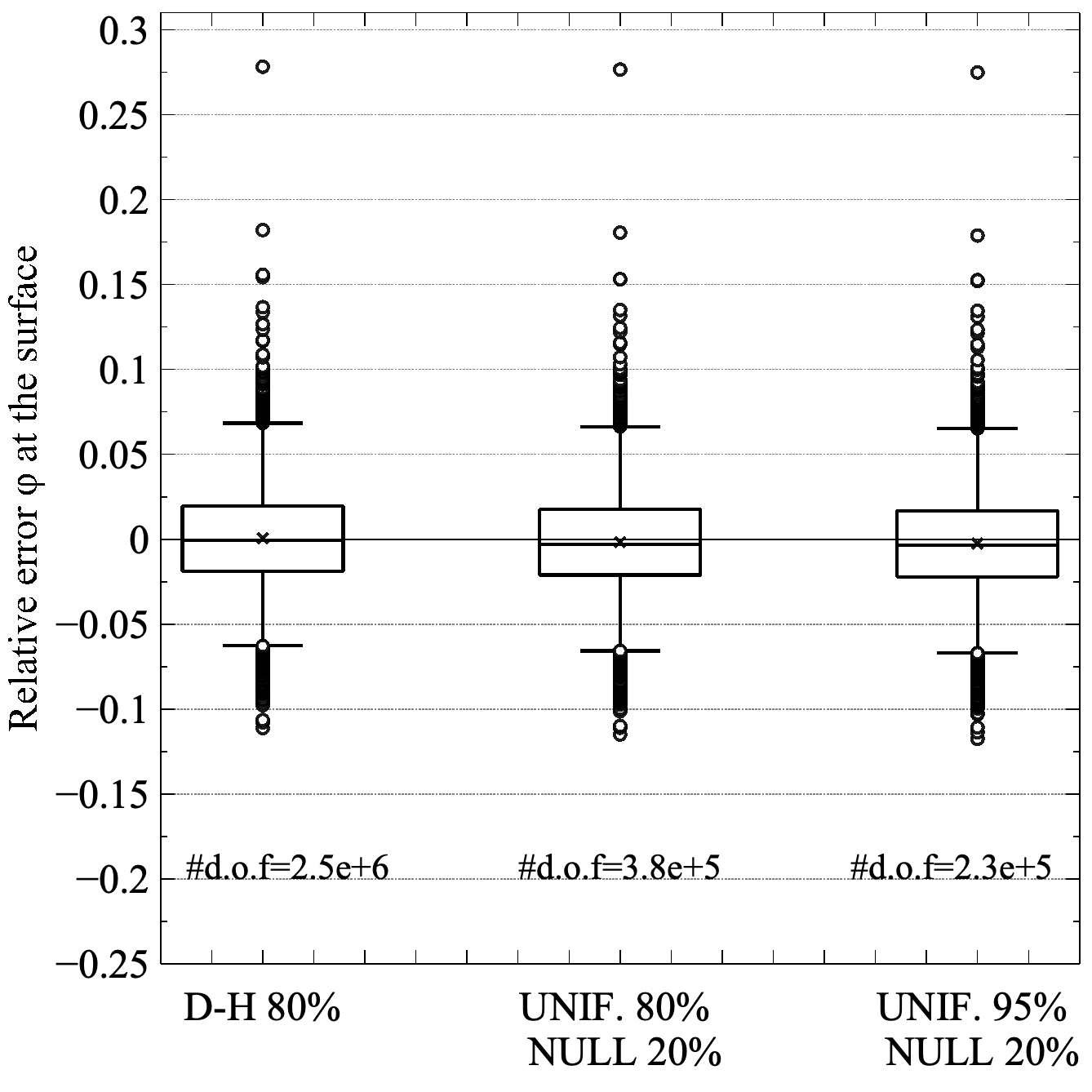}
    \caption{Box plot of relative error of potential at the surface compared with analytical results on the 30-spheres system using different BCs. D-H/Coulombic BCs on a cubic grid with $80\%$ perfil are used in the leftmost column. In the second and third, parallelepipedal uniform grid (perfil of $80\%$ and $95\%$, respectively) followed by a de-refined one until $20\%$ are used. The reduction in d.o.f. from the first and the second columns is due to the aspect ratio of the system, which permits a further computational saving due to the parallelepipedal shape of the uniform grid.}
    \label{fig:conf_30_sfere_BC}
\end{figure}

We then show the results obtained by leaving a uniform grid resolution on a region of parallelepipedal shape around the solute and performing a grid de-refinement until a much larger computational domain is reached. This technique allows us to bring the boundaries of the computational domain at a distance where null Dirichlet BCs become accurate irrespective of the system-specific geometry. This is obtained without a significant increment of the number of degrees of freedom, which may even get reduced for highly non-globular solutes relative to standard cubic computational domains (see Fig.~\ref{fig:deraffinamento}). For the 30-spheres system, we compare the use of D-H/Coulombic BCs at a perfil of $80\%$ with null BCs, applying de-refinement from 80\% to 20\% (or smaller) and from 95\% to 20\% (or smaller). From the results, shown in Fig.~\ref{fig:conf_30_sfere_BC}, we can see excellent results for null BCs when we de-refine until $20\%$ perfil.
Moreover, adopting a parallelepipedal rather than cubic uniform grid and possibly going from a perfil of $80\%$, as commonly done in PB calculations, to that of $95\%$ may compensate for the d.o.f. increment due to those located in the solvent, in the de-refinement region. Overall, this appears to be a convenient solution in terms of accuracy and computational cost.
% As shown in Table \ref{tabella con valori BC}, this method improves the quality of the boundary conditions while reducing the degrees of freedom.

\subsection{Real biomolecular systems}
\paragraph{Binding energy calculation}

We now assess our computational solution by analyzing its convergence wrt grid resolution on real biomolecular systems. We take six different, representative complexes from the dataset proposed by Fenley and co-workers \cite{Fenley-dataset} and observe the convergence of the binding free energy ($\Delta\Delta G$) and the total energy wrt the resolution. $\Delta\Delta G$ is calculated~as:
\begin{equation}
    \Delta\Delta G = \Delta G_c - \Delta G_1 -\Delta G_2
\end{equation}
where $\Delta G$ is the total electrostatic energy as defined in Eq. \eqref{eq:energyPart}, and the subscripts $c$, $1$, and $2$ correspond to the complex and the unbound components, respectively. To measure the convergence rate, we set as reference the results obtained when the grid spacing is 0.2~\AA. Moreover, we compare our results with different Poisson-Boltzmann solvers, including CPB \cite{CPB}, MIBPB \cite{MIBPB}, DelPhi \cite{DELPHI}, PBSA \cite{PBSA}, APBS \cite{APBS} and FEM-BEM \cite{FEM-BEM}, by taking their results from the works \cite{nguyen2017accurate} and~\cite{FEM-BEM}.  \\
We set a 90$\%$ perfil uniform grid and de-refine until we reach about the 20$\%$ perfil. For each resolution, we randomly displace the solute’s centroid 30 times within a cube of half the grid spacing.

\begin{figure}
    \centering
    \includegraphics[width=1\linewidth]{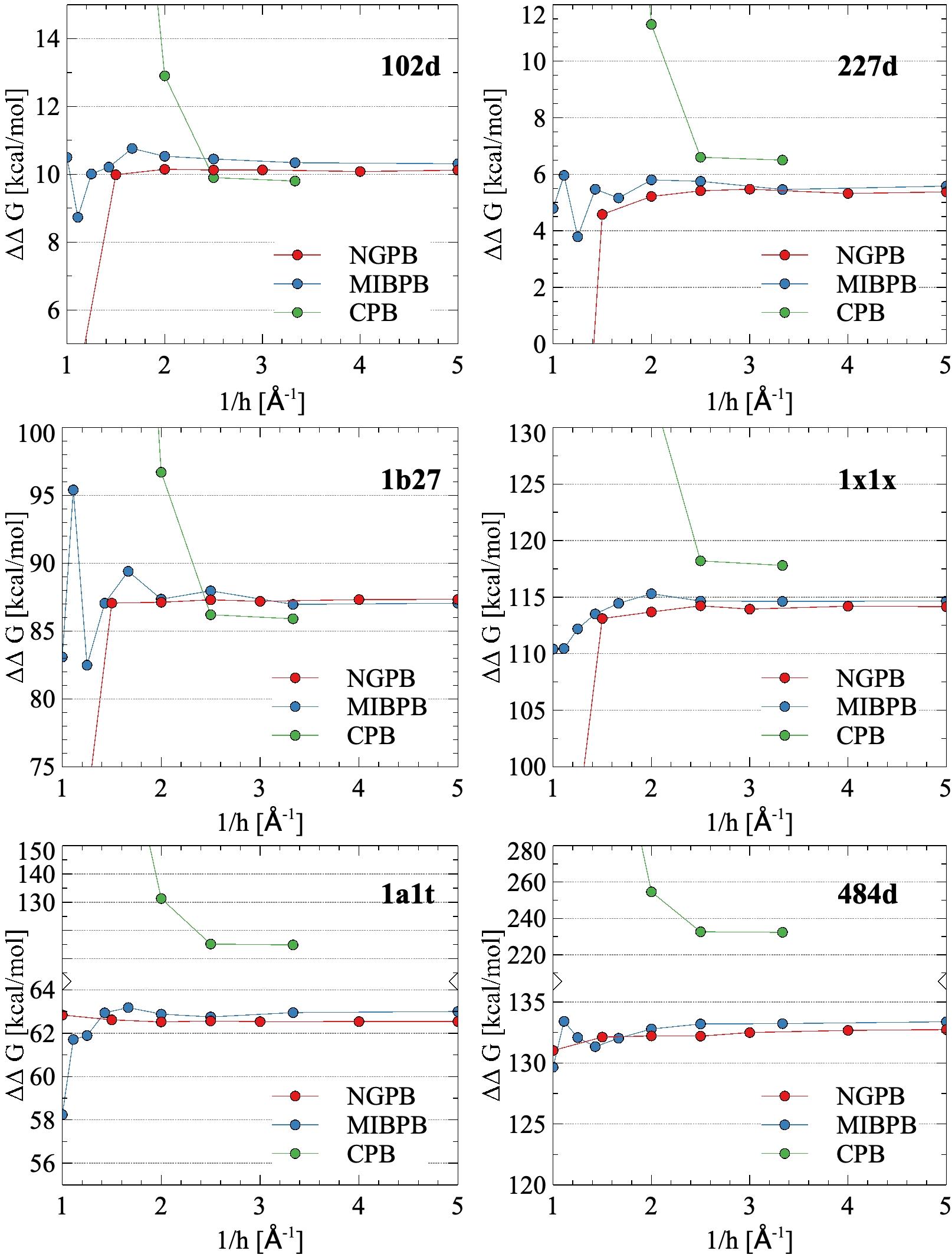}
    \caption{Convergence profile for $\Delta\Delta G$ and total free energy w.r.t. the finest mesh for NGPB, MIBPB and CPB.}
    \label{fig:converegence}
\end{figure}

\begin{figure}
    \centering
    \includegraphics[width=0.6\linewidth]{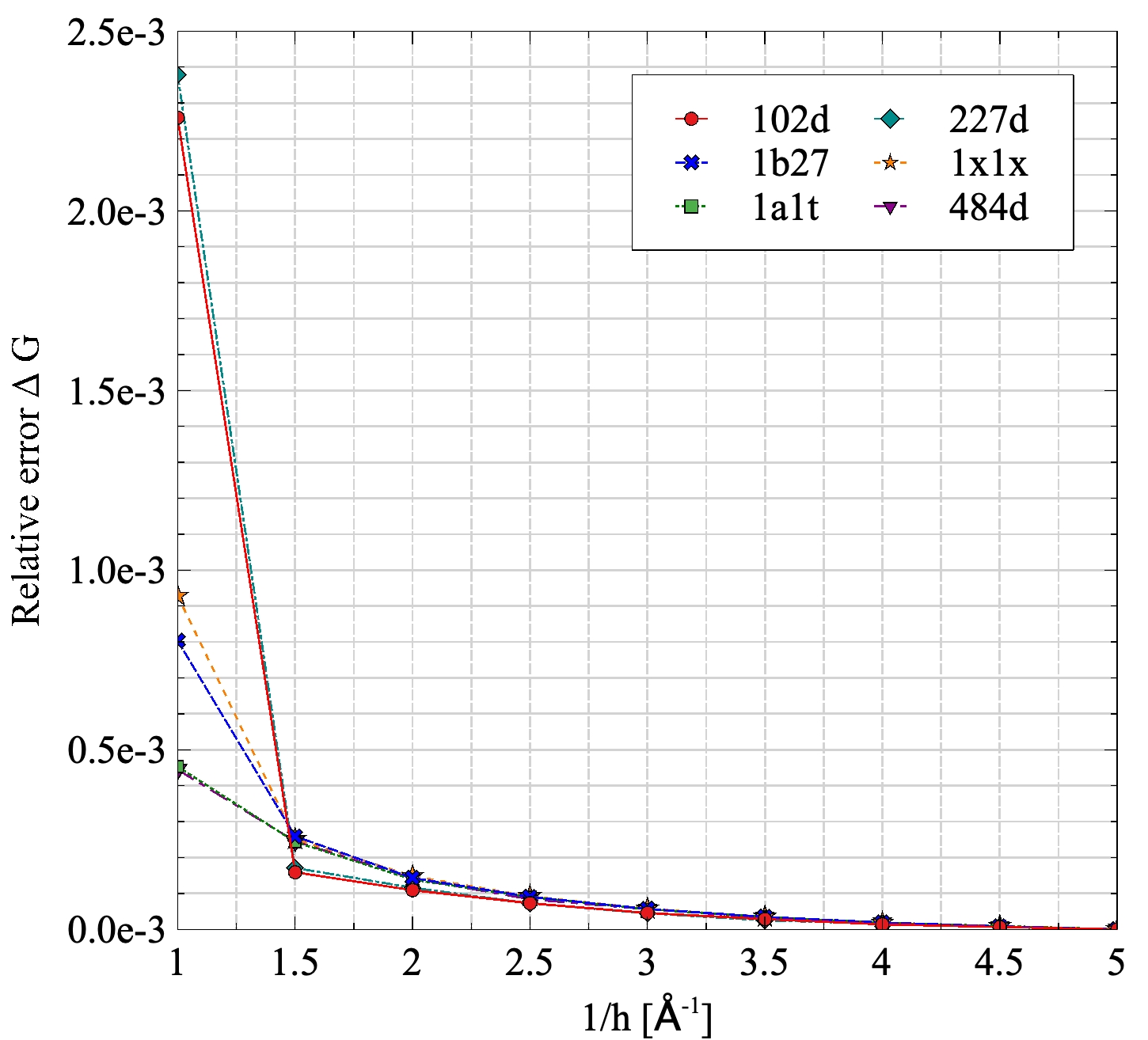}
    \caption{Convergence profile for total free energy w.r.t. the finest mesh for NGPB.}
    \label{fig:conv_energy}
\end{figure}

In Table \ref{tab:marcia_results}, we list the mean $\Delta\Delta G$ and the total energy, along with their respective standard deviations. Using data for CPB \cite{Fenley-dataset} and MIBPB \cite{nguyen2017accurate}, Fig. \ref{fig:converegence} shows the convergence plots for these solvers alongside those of NGPB. As one can see, we obtained satisfactory convergence at a grid spacing of 0.67~\AA~for $\Delta\Delta G$. As per the total energy, see Fig.~\ref{fig:conv_energy}, NGPB results are already quite accurate, starting from a grid spacing of 1~\AA. For calculating $\Delta\Delta G$, CPB reaches good convergence values for grid spacing values $\leq 0.4$~\AA, while MIBPB has better control over the results at larger spacings and a less rapid convergence in the low spacing range. It is worth noting that both local refinement around the surface, as done in CPB, and second-order accuracy enforcement, as done in MIBPB, entail a quite significant computational cost. The results for individual binding energies at the finest mesh for different solvers, also taken from \cite{Fenley-dataset,nguyen2017accurate}, are compared with those of NGPB.\\
Interestingly, while there can be quite significant differences among the results of different solvers and considering that there is no ground truth for these realistic systems, we can however observe that the results of NGPB are always close to those of MIBPB, which is supposedly among the most accurate solvers since it enforces second-order accuracy.\footnote{New results for the complete binding benchmark dataset show excellent agreement between CPB binding energies and those of MIBPB and NGPB (Marcia O. Fenley, private communication).} It is also important to note that the ionic component of the energy, which is typically negligible compared to other energy components, gains a more significant relative contribution to the binding energy when the systems involved are highly charged. As illustrated in Table \ref{tab:ddG_ionic}, we decompose the binding energy into its components, expressed as $\Delta\Delta G = \Delta\Delta G_{coul} + \Delta\Delta G_{pol} + \Delta\Delta G_{ion}$, and examine the relative significance of the ionic component ($\Delta\Delta G_{ion}$) in the total binding energy.
It is plausible that the observed discrepancies among some of the considered solvers, which are larger for more highly charged systems, are due to the specific ways used to calculate this term.

\begin{table}
    \centering
    \begin{tabular*}{0.61\columnwidth}{c|r|c}
    \hline
    Complex & Total charge $[e]$ & $\Delta \Delta G_{ion} /\Delta \Delta G $ \\
    \hline
    102d & -20 & 0.90 \\
    227d & -20 & 1.71 \\
    1b27 & -4  & 0.03 \\
    1x1x & -3  & 0.02 \\
    1a1t & 8   & 0.59 \\
    484d & -16 & 0.35 \\
    \hline
    \end{tabular*}%
    \caption{Relative importance of the ionic part of the binding energy wrt $\Delta\Delta G$ for each complex.}
    \label{tab:ddG_ionic}
\end{table}

\begin{figure}
    \centering
    \includegraphics[width=1\linewidth]{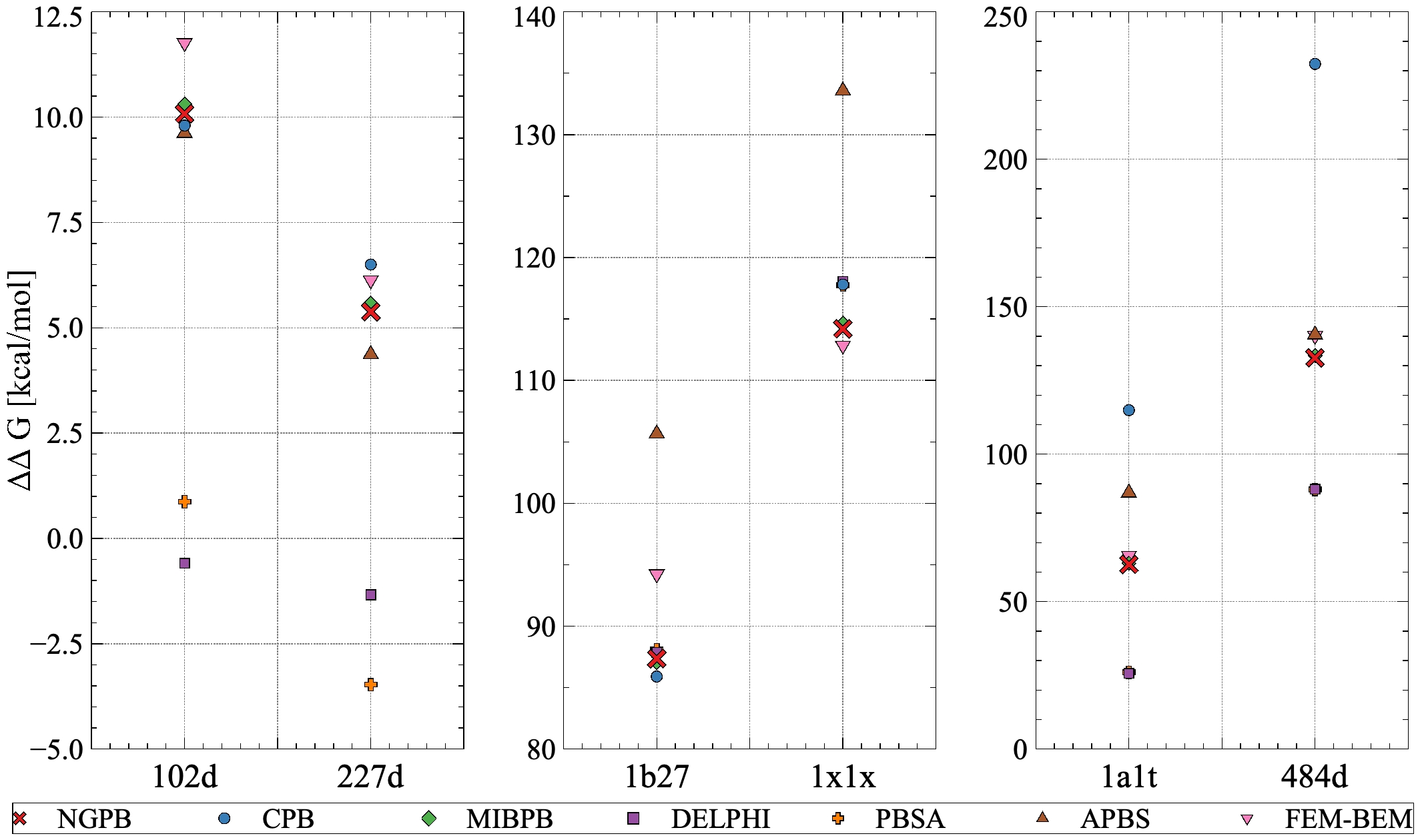}
    \caption{Binding energies for different solvers at $0.2$~\AA~grid spacing.}
    \label{fig:ddg_solvers}
\end{figure}

\begin{table}[!h]
    \centering
    \begin{tabular}{c|c c| c c| c c}
         &
        \multicolumn{2}{c|}{102d} & \multicolumn{2}{c|}{1b27} & \multicolumn{2}{c}{1a1t} \\[5pt]
        h[\AA] & $\Delta\Delta G$ & $\sigma$ & 
        $\Delta\Delta G$ & $\sigma$ & 
        $\Delta\Delta G$ & $\sigma$  \\[5pt]
       \hline
     1.00 &
     11.16  &   28.10  & 54.49  &   12.44 & 63.17  &    2.30 \\[1.5pt]  
    
     \hline
     0.67 & 
     9.87   &   1.34  & 86.71  &    2.41 & 62.82  &    1.09 \\[1.5pt] 
   
     \hline
     0.50 &
     10.01   &   0.42  & 87.16  &    0.99 & 62.35  &    0.68 \\[1.5pt] 
          
     \hline
     0.40 &
     10.24   &   0.43  & 87.31  &    0.72 & 62.79  &    0.50 \\[1.5pt] 

     \hline
     0.33 &
     10.08   &   0.42  & 87.34  &    0.34 & 62.54  &    0.26 \\[1.5pt] 

     \hline
     0.25 &
     10.08   &   0.28  & 87.30  &    0.22 & 62.51  &    0.23 \\[1.5pt] 
     
     \hline
     0.20 &
     10.08   &   0.20  & 87.35  &    0.17 & 62.59  &    0.19  \\[3.5pt] 
     
     \hline\\
     & $\Delta G$ & $\sigma$ & 
        $\Delta G$ & $\sigma$ & 
        $\Delta G$ & $\sigma$  \\[5pt]
       \hline
     1.00 &
     -25612.73  &   21.94 & -64947.35  &    6.20& -41003.42 &     1.92 \\[1.5pt]  
    
     \hline
     0.67 & 
     -25550.94  &    1.13 & -64982.73  &    1.49 & -41012.02  &    0.80 \\[1.5pt] 
   
     \hline
     0.50 &
     -25552.22  &    0.34 & -64990.27  &    0.79 & -41016.33  &    0.37\\[1.5pt] 
          
     \hline
     0.40 &
     -25553.14  &    0.29 & -64993.69  &    0.43 & -41018.32  &    0.29 \\[1.5pt] 

     \hline
     0.33 &
     -25553.83  &    0.30 & -64995.87  &    0.19 & -41019.71  &    0.11 \\[1.5pt] 

     \hline
     0.25 &
     -25554.65  &    0.18 & -64998.33  &    0.13 & -41021.29  &    0.15 \\[1.5pt] 
     
     \hline
     0.20 &
     -25555.00  &    0.14 & -64999.58  &    0.08 &  -41022.01  &    0.09\\[1.5pt]
       
    \hline
     \hline\\
         &
        \multicolumn{2}{c|}{227d} & \multicolumn{2}{c|}{1x1x} & \multicolumn{2}{c}{484d} \\[5pt]
        h[\AA] & $\Delta\Delta G$ & $\sigma$ & 
        $\Delta\Delta G$ & $\sigma$ & 
        $\Delta\Delta G$ & $\sigma$  \\[5pt]
       \hline
     1.00 &
     -17.30  &  26.39  & 74.41  &  13.91 & 130.84  &   4.09 \\[1.5pt]  
    
     \hline
     0.67 & 
     4.58   &  1.82  & 112.77  &   1.84 & 131.93 &    0.93 \\[1.5pt] 
   
     \hline
     0.50 &
     5.43   &  1.02  & 113.87  &   0.89 & 132.16 &    0.57 \\[1.5pt] 
          
     \hline
     0.40 &
     5.43   &  0.53  & 114.12  &   0.63 & 132.00 &    0.52 \\[1.5pt] 

     \hline
     0.33 &
     5.43   &  0.27  & 114.05  &   0.51 & 132.37 &    0.46 \\[1.5pt] 

     \hline
     0.25 &
     5.47  &   0.25  & 114.09  &   0.44 & 132.60 &   0.30 \\[1.5pt] 
     
     \hline
     0.20 &
     5.38  &   0.18  & 114.19 &    0.22 & 132.63 &   0.30  \\[3.5pt] 
     
     \hline\\
     & $\Delta G$ & $\sigma$ & 
        $\Delta G$ & $\sigma$ & 
        $\Delta G$ & $\sigma$  \\[5pt]
       \hline
     1.00 &
     -27099.69  &   22.64 & -64663.93  &    5.28 & -39427.15 &    2.91 \\[1.5pt]  
    
     \hline
     0.67 & 
     -27030.78  &   1.43 & -64707.76  &   1.08 & -39435.01    &  0.66 \\[1.5pt] 
   
     \hline
     0.50 &
     -27032.26  &   0.45 & -64714.43 &     0.61 & -39439.08 &     0.48\\[1.5pt] 
          
     \hline
     0.40 &
     -27033.44  &   0.35 & -64718.07 &    0.59 & -39441.38  &    0.34 \\[1.5pt] 

     \hline
     0.33 &
     -27034.18  &   0.18 & -64720.31   &   0.44 & -39442.46 &     0.24 \\[1.5pt] 

     \hline
     0.25 &
     -27034.96  &   0.18 & -64722.83  &   0.268 & -39443.98   &   0.18 \\[1.5pt] 
     
     \hline
     0.20 &
     -27035.40  &    0.14 & -64724.03  &   0.17 &  --39444.68 &    0.09
       
    \end{tabular}
    \caption{Convergence results for electrostatic free energies and $\Delta\Delta G$.}
    \label{tab:marcia_results}
\end{table}

\begin{figure}
    \centering
    \includegraphics[width=\textwidth]
    %{new_images/h1n1_screen_mid.png}
    %{new_images/h1n1_screen_half_hd.png}
    {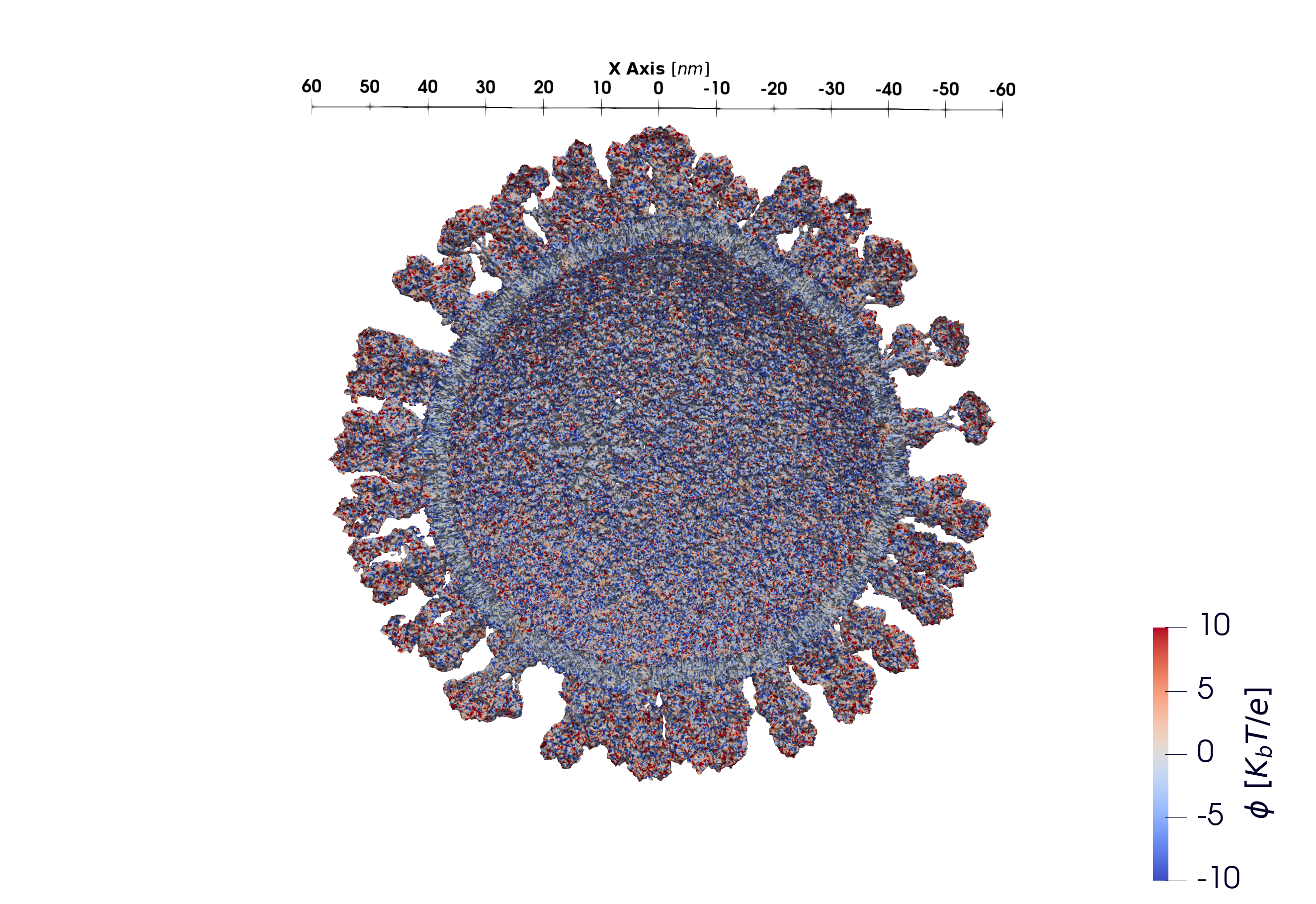}
    \caption{Half structure of the H1N1 swine virus having the molecular surface colored based on the local electrostatic potential.}
    \label{fig:h1n1_tot}
\end{figure}

\begin{figure}
    \centering
\includegraphics[width=0.5\textwidth]{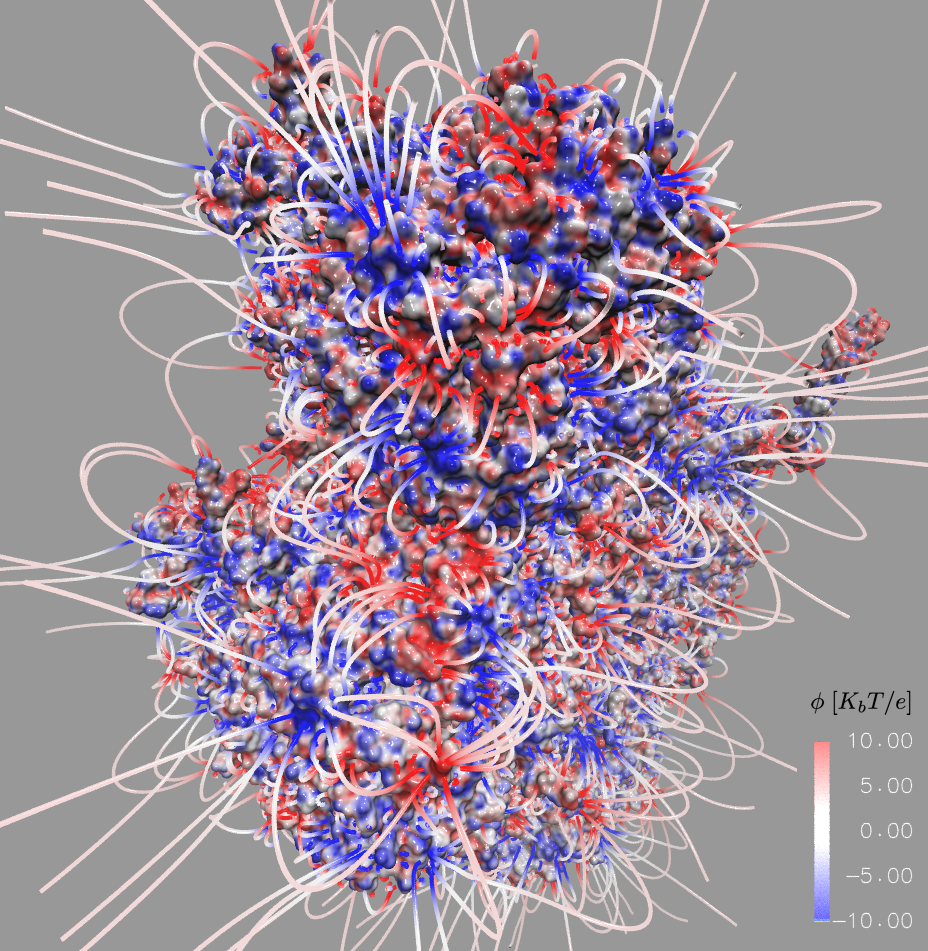}
    \caption{ Human adenovirus structure (pdb id 1VSZ) with the molecular surface colored based on the local electrostatic potential and showing main electric field lines.}
    \label{fig:1vsz}
\end{figure}

\paragraph{Application to large systems}
The most recent experimental techniques are providing us with structures at atomic or near-atomic resolution of unprecedented size. The computational challenges in solving such large systems' electrostatics are not trivial. Here, NextGenPB was challenged to calculate the electrostatic potential of two viral structures: the swine virus H1N1 capsid and and the human adenovirus (pdb id 1VSZ). For the H1N1 capsid, a grid resolution of 2 Å was used while for the 1VSZ system, a finer resolution of 0.5 Å was employed, to better capture the local nuances of the electrostatic potential, which is particularly important for understanding the molecular interactions. The resulting electrostatic potentials were plotted on the molecular surfaces of both systems, providing visual insights into the charge distribution and potential hotspots. These results are illustrated in Fig.~\ref{fig:h1n1_tot} for H1N1 and Fig.~\ref{fig:1vsz} for 1VSZ.

\newpage

%% file: Conclusions.tex
\section{Discussion and Conclusions}
\label{sec:conclusions}

Here we present a new solving and analyzing framework for the Poisson-Boltzmann equation where we leverage the availability of a very accurate description of the molecular surface, as provided by the NanoShaper software, by employing some analytical derivations that are used to customize the Cartesian FEM used to discretize and then solve the equation. 
The analytical contributions and corrections were used to make the expressions of the stiffness and mass matrices more accurate and to accelerate the calculation of the different components of the solvated system's electrostatic energy.

Interestingly, we also constructed an analytical benchmark consisting of combinations of non-overlapping spheres where numerical PB methods can be challenged. The availability of analytical results concerning more realistic systems is of paramount importance for making educated assessments of the solvers' quality.

On the numerical side, we use a Cartesian grid that can be locally refined, if needed. In our experience, however, the best trade-off between accuracy and computational cost for biomolecular systems consists of adopting a uniform grid of $h = 0.5$~\AA, tightly enclosing the system, and de-refining it (via de-duplication) until a global percentage of filling between $15$ and $20\%$ is reached. At the boundaries of the larger domain, homogeneous Dirichlet boundary conditions are applied. Together with our treatment of the dielectric discontinuity, this allows a good quality determination of the different energy contributions and surface potential while limiting the overall number of degrees of freedom and, therefore, the computational cost. 

Our analysis corroborates that, when calculated via the electric displacement flux method, the polarization energy term is quite robust concerning grid artifacts and boundary condition accuracy. In contrast, the value of the potential at the surface and, by consequence, that of the ionic contribution are more sensitive to both aspects.

We validate our new approach first against the ground truth represented by our analytical expressions, which, by separately providing the different energy terms, allow for a detailed analysis of the accuracy of the results. Then, we consider real biomolecular systems for which no analytical expression is available. Our figure of merit, in this case, is the rate of convergence of the total energy wrt grid resolution for different relative mutual placements wrt the grid. We use existing simulation data to compare our performance with that of some widely used codes.

Overall, the performance of NextGenPB is remarkable: in terms of accuracy, it reproduces analytically-derived electrostatic potentials at the molecular surface and energies with a relative error which is at least one order of magnitude smaller than that of the considered alternatives at the same grid resolution. 
On realistic systems, it quickly converges to the target value, and it yields results very close to those of MIBPB, which is expected to be the most accurate alternative since it enforces second-order accuracy.
Interestingly, the capability to accurately calculate the ionic contribution allowed highlighting its relevance in estimating the $\Delta\Delta G$. Indeed, while the ionic contribution is a small fraction of the others in the $\Delta G$, it can become prominent, due to cancellations of the other terms, if the systems are highly charged, as it occurs with nucleic acids, when the $\Delta\Delta G$ are derived.

In regard to computational efficiency, the better description of the dielectric discontinuity at almost no cost allows for better accuracy without the need to perform local refinements or second-order accuracy enforcement, which are computationally costly. Moreover, the derivation of the ionic contribution as an integral over the molecular surface allows its separate determination in only one run, in contrast to what is done in other PB solvers, where two runs at different ionic strengths are often performed. 
Finally, using grid de-refinement to impose null Dirichlet boundary conditions at a large distance from the solute allows saving the calculation of D-H-like ones without compromising on accuracy and while still keeping limited the overall number of degrees of freedom.

Under these considerations, the proposed approach, which marries analytical calculations and established numerical approaches, proves to achieve noteworthy cost-effectiveness. It paves the way for accurate and accelerated derivations of electrostatic potentials on large datasets that can be used to feed, for instance, machine learning tools.
A prototypical version of NGPB was incorporated in the MCCE software to describe the protonation states of proteins \cite{MCCE2}, which yielded auspicious results.
The future outlook for this activity is to transfer as many as possible of these improvements to a full nonlinear PB solver. Moreover, we are working on letting NGPB inherit the ability of NanoShaper to automatically filter out the internal cavities that are smaller than a given threshold in volume, in order to give the user the freedom to decide where they want the high-dielectric constant to be assigned to.
As a general comment, it is conceivable that the approach of using local analytical solutions in the elements crossing a sharp boundary can be profitably applied to other PDEs. Indeed, models leading to finite discontinuities in the constitutive parameters are relatively common in many fields, for instance, in models generated via Computer-Assisted Design.

\section{Acknowledgments}
\label{sec:Acknowledgements}

We acknowledge the financial support from the European 
Union - NextGenerationEU and the Ministry of University and Research (MUR), 
National Recovery and Resilience Plan (NRRP): 
Research program CN00000013 “National Centre for HPC, 
Big Data and Quantum Computing”, funded by the D.D. n.1031 del 17.06.2022 and Mission 4, 
Component 2, Investment 1.4 - Avviso “Centri Nazionali” - D.D. n. 3138, 16 December 2021; and 
PNRR MUR Project PE0000013 "Future Artificial Intelligence Research (FAIR)", CUP J53C22003010006.

The simulations discussed in this work have been performed on the HPC Cluster of the Department of Mathematics of Politecnico di Milano which was funded by MUR grant Dipartimento di Eccellenza 2023-2027, on the Franklin supercomputer of IIT, and on the CINECA supercomputers, thanks to EUROHPC and ISCRA initiatives.

The authors would like to thank Prof. C. Burstedde for fruitful discussion about the implementation of an efficient algorithm for depositing point charges on octree grids. They thank also Prof. M. O. Fenley for discussing the analysis of the results, and Dr. J. Chen, for providing a customized MIBPB code for performing our tests.

A special thanks goes to Martina Politi, whose master thesis at Politecnico di Milano~\cite{martina} provided the first proof--of--concept for the development of NextGenPB.

C. de Falco is a member of the Gruppo Nazionale  Calcolo Scientifico-Istituto Nazionale di Alta Matematica  (GNCS-INdAM).

V. Di Florio is a member of Gruppo Nazionale di Fisica Matematica (GNFM) of Istituto Nazionale di Alta Matematica (INdAM).

%A, B, C are part of ...\\

%% file: Appendix.tex
\section{Application of Green's identities to  partition the internal potential}
\label{appendix_internal_potential}
\noindent
Let's write Eq. \eqref{eq:LPBE} in operatorial form:
\begin{equation}
    L[\phi] = \rho^f \; ,
\label{eq:LPB_equation}
\end{equation}
where  the linear operator $L$ is defined as:
\begin{equation}
    L = -\nabla \cdot \left( \varepsilon(\mathbf{r})\nabla(\cdot) \right)
        + c(\mathbf{r}) (\cdot) \qquad \mathbf{r}\in \text{$D$}\; .
\label{eq:LPB_operator1}
\end{equation}

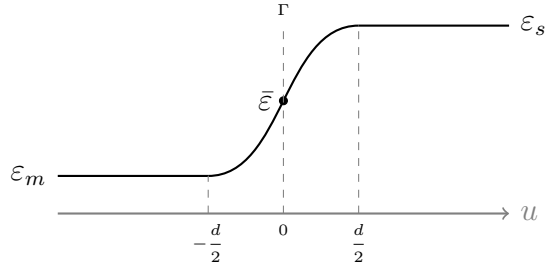
\begin{figure}[b!]
\centering
\begin{tikzpicture}
  \draw[black, thick] (0,0) node[left]{$\varepsilon_m$}  -- (2,0); 
  % \draw[black, thick] (2,0)  -- (4,2);
  % Tratto sigmoide con curva Bézier
  \draw[black, thick, smooth] (2,0) .. controls (3,0) and (3,2) .. (4,2);
  \draw[black, thick] (4,2)   -- (6,2) node[right]{$\varepsilon_s$};
  \filldraw[black] (3,1) circle (1.5pt) node[left]{$\Bar{\varepsilon}$};
  \draw[->,gray, thick] (0,-0.5)-- (6,-0.5) node[right]{$u$}; %AXIS
  \draw[gray, dashed] (2,-0.5) node[black, below]{\tiny $-\dfrac{d}{2}$}-- (2,0);
  \draw[gray, dashed] (4,-0.5) node[black, below]{\tiny $\dfrac{d}{2}$}-- (4,2);
  \draw[gray, dashed] (3,-0.5) node[black, below]{\tiny $0$} -- 
                       (3,2) node[black,above]{\tiny $\Gamma$};
\end{tikzpicture}
\caption{Schematic representation of the approximated variation of~$\varepsilon$.}
\label{fig:lin_eps}
\end{figure}

A straightforward application of the second Green's identity to the operator $L$ is not possible, since $\varepsilon(\mathbf{r})$ is not continuous and, consequently, the electrostatic potential is not differentiable at the surface $\Gamma$. Therefore, we will attempt a non rigorous derivation, where the gradient of $\varepsilon$ is considered in a distributional sense.
More in detail, we consider $\varepsilon(\mathbf{r})$, which is a step function with the discontinuity located on $\Gamma$, as a limit for $d \to 0^+$ of the following function:
\begin{equation}
    \varepsilon_d(\mathbf{r}) = \dfrac{\varepsilon_s - \varepsilon_m}{2} \left [\operatorname{erf}\left(\dfrac{u(\mathbf{r})}{d}\right) + 1 \right] + \varepsilon_m\; .
\label{eq:approximant}
\end{equation}
For the definition of $u$, we start by considering for each $\mathbf{r}$ near the surface, its orthogonal projection on $\Gamma$, namely $\mathbf{r}_{\Gamma} \in \Gamma$, and its direction $\mathbf{n}(\mathbf{r}_{\Gamma})$. 
We define $u$ as the oriented distance from the surface itself: $u(\mathbf{r}) = (\mathbf{r}-\mathbf{r}_{\Gamma}) \cdot \mathbf{n}(\mathbf{r}_{\Gamma})$.
In this approximation, $\varepsilon_d(u=0) = \Bar{\varepsilon} = \dfrac{\varepsilon_m +\varepsilon_s}{2}$. A schematic general representation of $\varepsilon_d$ is shown in Fig.~\ref{fig:lin_eps}. It is interesting to note that numerical FEM-BEM-coupled modelling with MS represented as a continuous interface (similarly to \eqref{eq:approximant} with non-zero $d$) and the corresponding impact of such diffused interfaces on solvation and binding energy, were very recently treated in~\cite{FEM-BEM_diffuse_interface}. In addition to the error function $\operatorname{erf}$, the work \cite{FEM-BEM_diffuse_interface} has benchmarked other sigmoidal functions (like $\tanh$) to couple the internal and external regions.

Let's now consider the gradient of $\varepsilon_d$, it is aligned with the normal at the surface and takes the following form:

\begin{equation}
    \nabla \varepsilon_d(\mathbf{r}) = \dfrac{\mathrm{d} \varepsilon_d(u)}{\mathrm{d} u} \nabla u  = (\varepsilon_s-\varepsilon_m)\mathcal{N}(0,\dfrac{d^2}{2})\mathbf{n}(\mathbf{r}_{\Gamma}) 
\end{equation}
(here the symbol $\mathcal N(0,d^2/2)$ indicates the normal distribution density function with zero mean and variance~$d^2/2$).

In the $d\rightarrow 0$ limit, we obtain the gradient expression:
\begin{equation}
\label{grad_epsilon_surf}
    \nabla \varepsilon(\mathbf{r}) = (\varepsilon_s-\varepsilon_m)\delta((\mathbf{r}-\mathbf{r}_{\Gamma}) \cdot\mathbf{n}(\mathbf{r}_{\Gamma})) \mathbf{n}(\mathbf{r}_{\Gamma}) \qquad  \mathbf{r}_{\Gamma}\in \Gamma \; .
\end{equation}
In what follows, we assume that the limits for $d \to 0$ and the integral operators can be swapped. 
Similarly to what is done in the case of the second Green's identity, we consider the following identity:
\begin{equation}
    \int_{\Omega} \left\{L[v]w - L[w]v \right\} \dd V =
    \int_{\Omega} \nabla \cdot \left [ \varepsilon(\mathbf{r}) \left ( v \nabla w - w \nabla v \right) \right ] \dd V \;
   % \left[ \varepsilon(\Tilde{\mathbf{r}}) (\nabla w)_n v -
    %\varepsilon(\Tilde{\mathbf{r}}) (\nabla v)_n w\right] \dd S \; ,
\label{eq:1_green_id}
\end{equation}
where $\Omega \subseteq D$, $w$ and $v$ are suitable test functions, $\Tilde{\mathbf{r}}$ is the integration variable. \\

Let us now instantiate Eq.~\eqref{eq:1_green_id} by choosing $\Omega = \Omega_m$, $w(\Tilde{\mathbf{r}}) = \phi (\Tilde{\mathbf{r}})$ as the solution of Eq.~\eqref{eq:LPB_equation}, and $v(\Tilde{\mathbf{r}}) = \dfrac{q}{4\pi \varepsilon_m \| \Tilde{\mathbf{r}}-\mathbf{r} \|}$ representing the point charge potential centered at the observation point $\mathbf{r}\in \Omega_m$. The left-hand side (LHS) of Eq.~\eqref{eq:1_green_id} then takes on the following form:
\begin{equation}
\begin{split}
   \int_{\Omega_m} & \left\{L[v(\Tilde{\mathbf{r}})]\phi (\Tilde{\mathbf{r}}) - L[\phi (\Tilde{\mathbf{r}})]v(\Tilde{\mathbf{r}}) \right\} \dd V = \int_{\Omega_m} 
      \biggl\{ q\phi(\Tilde{\mathbf{r}})
      \biggl[
        \dfrac{c(\Tilde{\mathbf{r}})}{4\pi \varepsilon_m \| \Tilde{\mathbf{r}}-\mathbf{r} \|} + \\ &+
        \dfrac{\varepsilon(\Tilde{\mathbf{r}})}{\varepsilon_m}\delta(\Tilde{\mathbf{r}}-\mathbf{r}) +
         \dfrac{\nabla(\varepsilon(\Tilde{\mathbf{r}}))\cdot(\Tilde{\mathbf{r}}-\mathbf{r})}{4\pi\varepsilon_m \| \Tilde{\mathbf{r}}-\mathbf{r} \|^3} 
      \biggr]  -
       \dfrac{\rho^f q}{4\pi \varepsilon_m \| \Tilde{\mathbf{r}}-\mathbf{r} \|} \biggr\} \dd V  \; .
\end{split}
\label{eq:LHS_green}
\end{equation}
%   &=\int_{\Omega} \left\{
%      \phi(\Tilde{\mathbf{r}}) 
%      \left[ 
%        \dfrac{c(\Tilde{\mathbf{r}})q}{4\pi \varepsilon_m \| \Tilde{\mathbf{r}}-\mathbf{r} \|} -
%        \nabla\left(
%  \varepsilon(\Tilde{\mathbf{r}})\nabla\dfrac{q}{4\pi\varepsilon_m  \| \Tilde{\mathbf{r}}-\mathbf{r} \|} \right)
%      \right] -
%      \rho^f \dfrac{q}{4\pi \varepsilon_m \| \Tilde{\mathbf{r}}-        \mathbf{r} \|}
%    \right\} \dd \Tilde{\mathbf{r}} \\
In $\Omega_m$, $c(\Tilde{\mathbf{r}}) =0$ and $\varepsilon(\mathbf{r}) = \varepsilon_m$. 
Relation \eqref{eq:LHS_green} then becomes:
\begin{equation}
\begin{aligned}
\int_{\Omega_m} & \{ L {} [v(\Tilde{\mathbf{r}})]\phi (\Tilde{\mathbf{r}}) - L[\phi (\Tilde{\mathbf{r}})]v(\Tilde{\mathbf{r}}) \} \dd V =\\
    &=q\phi(\mathbf{r}) +
    \dfrac{q}{4\pi\varepsilon_m }\int_{\Omega_m} \phi(\Tilde{\mathbf{r}})  \dfrac{\nabla(\varepsilon(\Tilde{\mathbf{r}}))\cdot(\Tilde{\mathbf{r}}-\mathbf{r})}{\| \Tilde{\mathbf{r}}-\mathbf{r} \|^3} \dd V -        
      \dfrac{q}{4\pi\varepsilon_m }\int_{\Omega_m} \dfrac{\rho^f}{\|\Tilde{\mathbf{r}} -\mathbf{r}\|} \dd V =\\
    &=q\phi(\mathbf{r}) +
    \dfrac{q}{4\pi\varepsilon_m }\int_{\Omega_m} \phi(\Tilde{\mathbf{r}}) (\varepsilon_s-\varepsilon_m) \dfrac{\delta((\mathbf{r}-\mathbf{r}_{\Gamma}) \cdot\mathbf{n}(\mathbf{r}_{\Gamma})) \mathbf{n}(\mathbf{r}_{\Gamma})\cdot(\Tilde{\mathbf{r}}-\mathbf{r})}{\| \Tilde{\mathbf{r}}-\mathbf{r} \|^3} \dd V +\\ 
    &\qquad \qquad - \dfrac{q}{4\pi\varepsilon_m }\int_{\Omega_m} \dfrac{\rho^f}{\|\Tilde{\mathbf{r}} -\mathbf{r}\|} \dd V\; .
 \end{aligned}
\label{eq:A_in}
\end{equation}
\noindent
Thanks to the \emph{single layer} property of the Dirac delta function and considering that, along the limit for $d\to 0$ process, we are integrating only on the internal half of the layer enclosing the surface where $\varepsilon_d$ is changing, we can recast the following volume integral in a surface one: 
\begin{equation}
\begin{aligned}
    \int_{\Omega_m} \phi(\Tilde{\mathbf{r}}) (\varepsilon_s-\varepsilon_m) \dfrac{\delta((\mathbf{r}-\mathbf{r}_{\Gamma}) \cdot\mathbf{n}(\mathbf{r}_{\Gamma})) \mathbf{n}(\mathbf{r}_{\Gamma})\cdot(\Tilde{\mathbf{r}}-\mathbf{r})}{\| \Tilde{\mathbf{r}}-\mathbf{r} \|^3} \dd V =\\
    = \dfrac{\varepsilon_s -\varepsilon_m}{2}
       \int_{\partial \Omega_m}  \phi(\Tilde{\mathbf{r}}) \dfrac{(\Tilde{\mathbf{r}} - \mathbf{r}) \cdot \mathbf{n}(\Tilde{\mathbf{r}})}{\|\Tilde{\mathbf{r}} - \mathbf{r} \|^3}  \mathrm{d}S \; .
\end{aligned}
\label{eq:single_layer}
\end{equation}
\noindent
In the present case, $\partial \Omega_m \equiv \Gamma$ and $\mathbf{n}$
denotes the unit vector orthogonal to $\Gamma$ and oriented in the outward direction.\\
Let us now consider the right-hand side (RHS) of \eqref{eq:1_green_id} in our instantiation:
\begin{equation}
\begin{split}
\int_{\Omega_m} \nabla \cdot & [\varepsilon(\Tilde{\mathbf{r}}) v(\Tilde{\mathbf{r}}) \nabla \phi (\Tilde{\mathbf{r}}) - \varepsilon(\Tilde{\mathbf{r}})\phi (\Tilde{\mathbf{r}}) \nabla v(\Tilde{\mathbf{r}})] \dd V =\\
    &= -\int_{\Omega_m}\nabla \cdot [ v(\Tilde{\mathbf{r}}) \mathbf{D}(\Tilde{\mathbf{r}})]\dd V 
    +\int_{\Omega_m} \nabla \cdot [\varepsilon(\Tilde{\mathbf{r}}) \phi (\Tilde{\mathbf{r}}) \dfrac{q(\Tilde{\mathbf{r}}-\mathbf{r})}{4\pi \varepsilon_m \| \Tilde{\mathbf{r}}-\mathbf{r} \|^3} ] \dd V =\\  
    &= -q\int_{\Gamma} \dfrac{\mathbf{D}(\Tilde{\mathbf{r}}) \cdot \mathbf{n}(\Tilde{\mathbf{r}})}{4\pi \varepsilon_m \| \Tilde{\mathbf{r}}-\mathbf{r} \|}  \dd S + 
       q\int_{\Gamma} \Bar{\varepsilon}\phi(\Tilde{\mathbf{r}}) \dfrac{(\Tilde{\mathbf{r}}-\mathbf{r}) \cdot \mathbf{n}(\Tilde{\mathbf{r}})}{4\pi \varepsilon_m \| \Tilde{\mathbf{r}}-\mathbf{r} \|^3}  \dd S \; ,
\end{split}
\label{eq:B_in}
\end{equation}
where $\mathbf D$ is the electric displacement vector and $(\nabla \cdot)_n = \nabla(\cdot) \cdot \mathbf{n}$. The application of the divergence theorem done in this latest derivations was a bit more natural in the integral involving $\mathbf{D}$ since its normal component is continuous, while the contribution to the integral given by the discontinuous tangential component is nullified by the scalar product with $\mathbf{n}$. In the second integral, in contrast, we have been forced to consider the limit of $\varepsilon_d$. 
Combining \eqref{eq:A_in} and \eqref{eq:B_in}, we can express the electrostatic potential in any point of $\Omega_m$ as follows:
\begin{equation*}
\begin{aligned}
\phi(\mathbf{r}) {} & = \int_{\Omega_m} \dfrac{\rho^f}  {4\pi\varepsilon_m\|\Tilde{\mathbf{r}} -\mathbf{r}\|} \dd V -
    \dfrac{\varepsilon_s -\varepsilon_m}{2}
    \int_{\Gamma}  \phi(\Tilde{\mathbf{r}}) \dfrac{(\Tilde{\mathbf{r}} - \mathbf{r}) \cdot \mathbf{n}(\Tilde{\mathbf{r}})}{4\pi \varepsilon_m \|\Tilde{\mathbf{r}} - \mathbf{r} \|^3}  \mathrm{d}S +\\
    & \qquad -\int_{\Gamma} \dfrac{\mathbf{D}(\Tilde{\mathbf{r}}) \cdot \mathbf{n}(\Tilde{\mathbf{r}})}{4\pi \varepsilon_m \| \Tilde{\mathbf{r}}-\mathbf{r} \|}  \dd S + 
    \int_{\Gamma} \Bar\varepsilon \phi(\Tilde{\mathbf{r}}) \dfrac{(\Tilde{\mathbf{r}}-\mathbf{r}) \cdot \mathbf{n}(\Tilde{\mathbf{r}})}{4\pi \varepsilon_m \| \Tilde{\mathbf{r}}-\mathbf{r} \|^3}  \dd S =\\
&=\int_{\Omega_m} \dfrac{\rho^f}{4\pi\varepsilon_m\|\Tilde{\mathbf{r}} -\mathbf{r}\|} \dd V -\int_{\Gamma} \dfrac{\mathbf{D}(\Tilde{\mathbf{r}}) \cdot \mathbf{n}(\Tilde{\mathbf{r}})}{4\pi \varepsilon_m \| \Tilde{\mathbf{r}}-\mathbf{r} \|}  \dd S +
    \int_{\Gamma}  \phi(\Tilde{\mathbf{r}}) \dfrac{(\Tilde{\mathbf{r}} - \mathbf{r}) \cdot \mathbf{n}(\Tilde{\mathbf{r}})}{4\pi \|\Tilde{\mathbf{r}} - \mathbf{r} \|^3}  \mathrm{d}S \;,
\end{aligned}  
% \label{eq:app_phi_in}
\end{equation*}
which actually yields Eq.~\eqref{eq:pot_in_partitioning}.

\section{Ionic potential: from volume to surface formulation}
\label{appendix_external_potential1}
\noindent
If we repeat the previous calculations by assuming $\Omega = \Omega_s$, and therefore accounting for the fact that in this region $\rho^f(\Tilde{\mathbf{r}}) = 0$ while $c(\Tilde{\mathbf{r}}) \neq 0$ (except for the Stern layer, if any), and by choosing as test function $v(\Tilde{\mathbf{r}}) = \dfrac{q}{4\pi\varepsilon_s \|\Tilde{\mathbf{r}}-\mathbf{r} \|}$, with $\mathbf{r} \in \Omega_m$, we obtain, for the LHS of Eq.~\eqref{eq:1_green_id}:
\begin{equation}
\begin{split}
\int_{\Omega_s} \left\{L[v(\Tilde{\mathbf{r}})]\phi (\Tilde{\mathbf{r}}) - L[\phi (\Tilde{\mathbf{r}})]v(\Tilde{\mathbf{r}}) \right\} \dd V
  =\int_{\Omega_s} 
    \dfrac{q\phi(\Tilde{\mathbf{r}})}{4\pi \varepsilon_s}
      \left[
        \dfrac{c(\Tilde{\mathbf{r}})}{\| \Tilde{\mathbf{r}}-\mathbf{r} \|} +    \dfrac{\nabla\varepsilon(\Tilde{\mathbf{r}})\cdot(\Tilde{\mathbf{r}}-\mathbf{r})}{\| \Tilde{\mathbf{r}}-\mathbf{r} \|^3} 
      \right] \dd V  \; .
\end{split}
\label{eq:A_out1}
\end{equation}
and for its right-hand-side:
\begin{equation}
\begin{split}
\int_{\Omega_s} \nabla \cdot & [\varepsilon(\Tilde{\mathbf{r}}) v(\Tilde{\mathbf{r}}) \nabla \phi (\Tilde{\mathbf{r}}) - \varepsilon(\Tilde{\mathbf{r}})\phi (\Tilde{\mathbf{r}}) \nabla v(\Tilde{\mathbf{r}})] \dd V =\\
    &= -q\int_{\Gamma \cup \Sigma} \dfrac{\mathbf{D}(\Tilde{\mathbf{r}}) \cdot \mathbf{m}(\Tilde{\mathbf{r}})}{4\pi \varepsilon_s \| \Tilde{\mathbf{r}}-\mathbf{r} \|}  \dd S + 
       q\int_{\Gamma \cup \Sigma} \Bar{\varepsilon}^\prime \phi(\Tilde{\mathbf{r}}) \dfrac{(\Tilde{\mathbf{r}}-\mathbf{r}) \cdot \mathbf{m}(\Tilde{\mathbf{r}})}{4\pi \varepsilon_s \| \Tilde{\mathbf{r}}-\mathbf{r} \|^3}  \dd S \; ,
\end{split}
\label{eq:B_out1}
\end{equation}
where $\Bar{\varepsilon}^\prime=\Bar{\varepsilon}$ on $\Gamma$ and $\Bar{\varepsilon}^\prime=\varepsilon_s$ on $\Sigma$, and $\mathbf{m}$ is the outward normal from $\Omega_s$; $\Gamma$ is the surface separating $\Omega_s$ from $\Omega_m$ 
and $\Sigma$ is the external surface enclosing~$\Omega_s$. If, as it is commonly assumed, the solvent region extends to infinity, it can be shown that the surface integrals over $\Sigma$ vanish. Consistently, the normal vector $\mathbf{m}$ on $\Gamma$ coincides with $-\mathbf{n}$.
If we now equate Eqs. \eqref{eq:A_out1} and \eqref{eq:B_out1}, and consider that $\rho^s(\Tilde{\mathbf{r}}) = - c(\Tilde{\mathbf{r}})\phi(\Tilde{\mathbf{r}})$, we get:
\begin{equation}
\int_{\Omega_s}
        \dfrac{\rho^s(\Tilde{\mathbf{r}})}{4\pi \varepsilon_s\| \Tilde{\mathbf{r}}-\mathbf{r} \|} \dd V
= -\int_{\Gamma} \dfrac{\mathbf{D}(\Tilde{\mathbf{r}}) \cdot \mathbf{n}(\Tilde{\mathbf{r}})}{4\pi \varepsilon_s \| \Tilde{\mathbf{r}}-\mathbf{r} \|} \dd S + 
\int_{\Gamma} \phi(\Tilde{\mathbf{r}}) \dfrac{(\Tilde{\mathbf{r}}-\mathbf{r}) \cdot \mathbf{n}(\Tilde{\mathbf{r}})}{4\pi\| \Tilde{\mathbf{r}}-\mathbf{r} \|^3}  \dd S \; .
\label{eq:id_green_out1}
\end{equation}
It is interesting to note that the LHS of the latter equation is the potential generated by the ions in solution screened by the polarizable solvent, evaluated at a point $\mathbf{r}$ inside the solute. This is the exact definition of the reaction potential coming from the ions in solution, that we call $\phi_{ion}$ (see \eqref{eq:pot_inside_final} and \eqref{eq:energyPart}). We get here a direct derivation of what has been found at the end of Sect.~\ref{sect.Partitioning}.

\section{Application of Green’s identities to partition the potential in the solvent}
\label{appendix_external_potential2}
\noindent
There is a third way to apply the same procedure seen in the two previous Appendices, to extract further interesting information. In this case, we again assume $\Omega = \Omega_s$, and account for the fact that in this region $\rho^f(\Tilde{\mathbf{r}}) = 0$ while $c(\Tilde{\mathbf{r}}) \neq 0$ (except that in the Stern layer, if any). In contrast to the previous applications, we choose as test function the D-H solution for a single charge in solution, with the observer location $\mathbf{r} \in \Omega_s$:
\begin{equation}
    v(\Tilde{\mathbf{r}}) = \frac{qe^{-k_D\|\Tilde{\mathbf{r}} -\mathbf{r}\|}}{4 \pi \varepsilon_s \|\Tilde{\mathbf{r}} -\mathbf{r}\|} \; .
\end{equation}
To simplify the derivation, we assume the distance of $\mathbf{r}$ from the surface $\Gamma$ is larger than any $d$ considered in the limiting process for $\varepsilon_d$ (see \ref{appendix_internal_potential}), so that we can always assume that $\varepsilon(\mathbf{r}) = \varepsilon_s$. 
The LHS of Eq.~\eqref{eq:1_green_id} yields 
\begin{equation}
\begin{split}
\int_{\Omega_s} & \left\{L[v(\Tilde{\mathbf{r}})]\phi(\Tilde{\mathbf{r}}) - L[\phi (\Tilde{\mathbf{r}})]v(\Tilde{\mathbf{r}}) \right\} \dd V =
  -q\int_{Stern~layer}
        \phi(\Tilde{\mathbf{r}})\dfrac{k_D^2 e^{-k_D \| \Tilde{\mathbf{r}}-\mathbf{r} \|}}{4\pi \| \Tilde{\mathbf{r}}-\mathbf{r} \|} \dd V +\\   
        &+q\int_{\Omega_s} \phi(\Tilde{\mathbf{r}})  \dfrac{\nabla \varepsilon(\Tilde{\mathbf{r}})\cdot(\Tilde{\mathbf{r}}-\mathbf{r})e^{-k_D\| \Tilde{\mathbf{r}}-\mathbf{r} \|}(1+k_D\| \Tilde{\mathbf{r}}-\mathbf{r} \|)}{4\pi \varepsilon_s \| \Tilde{\mathbf{r}}-\mathbf{r} \|^3} \dd V + q \phi(\mathbf{r}) \; ,
\end{split}
\label{eq:A_out2}
\end{equation}
where the term $q\phi(\mathbf{r})$ results from the convolution of $\phi(\Tilde{\mathbf{r}})$ with the Dirac delta arising from the PB operator, while the RHS of Eq.~\eqref{eq:1_green_id} yields:
\begin{equation}
\begin{split}
\int_{\Omega_s} \nabla \cdot & [\varepsilon(\Tilde{\mathbf{r}}) v(\Tilde{\mathbf{r}}) \nabla \phi (\Tilde{\mathbf{r}}) - \varepsilon(\Tilde{\mathbf{r}})\phi (\Tilde{\mathbf{r}}) \nabla v(\Tilde{\mathbf{r}})] \dd V =
-q\int_{\Gamma \cup \Sigma} \dfrac{e^{-k_D\| \Tilde{\mathbf{r}}-\mathbf{r} \|}\mathbf{D}(\Tilde{\mathbf{r}}) \cdot \mathbf{m}(\Tilde{\mathbf{r}})}{4\pi \varepsilon_s \| \Tilde{\mathbf{r}}-\mathbf{r} \|}  \dd S +\\
       &+ q\int_{\Gamma \cup \Sigma} \Bar{\varepsilon}^\prime \phi(\Tilde{\mathbf{r}}) \dfrac{e^{-k_D\| \Tilde{\mathbf{r}}-\mathbf{r} \|}(1+k_D\| \Tilde{\mathbf{r}}-\mathbf{r} \|)(\Tilde{\mathbf{r}}-\mathbf{r}) \cdot \mathbf{m}(\Tilde{\mathbf{r}})}{4\pi \varepsilon_s \| \Tilde{\mathbf{r}}-\mathbf{r} \|^3}  \dd S \; ,
\end{split}
\label{eq:B_out2}
\end{equation}
where $\Bar{\varepsilon}^\prime=\Bar{\varepsilon}$ on $\Gamma$ and $\Bar{\varepsilon}^\prime=\varepsilon_s$ on $\Sigma$, $\mathbf{m}$ is the outward normal from $\Omega_s$, $\Gamma$ is the surface separating $\Omega_s$ from $\Omega_m$ 
and $\Sigma$ is the external surface enclosing $\Omega_s$. If, as it is commonly assumed, the solvent region extends to infinity, it can be shown that the surface integrals over $\Sigma$ vanish. Consistently, the normal vector $\mathbf{m}$ on $\Gamma$ coincides with $-\mathbf{n}$.
If we now equate Eqs.~\eqref{eq:A_out2} and \eqref{eq:B_out2} and reorder the terms, we will finally get:
\begin{equation*}
\begin{split}
\phi(\mathbf{r}) = {}& 
    \int_{Stern~layer} \phi(\Tilde{\mathbf{r}})\dfrac{k_D^2 e^{-k_D\| \Tilde{\mathbf{r}}-\mathbf{r} \|}}{4\pi \| \Tilde{\mathbf{r}}-\mathbf{r} \|} \dd V
    + \int_{\Gamma} \dfrac{e^{-k_D\| \Tilde{\mathbf{r}}-\mathbf{r} \|}\mathbf{D}(\Tilde{\mathbf{r}}) \cdot \mathbf{n}(\Tilde{\mathbf{r}})}{4\pi \varepsilon_s \| \Tilde{\mathbf{r}}-\mathbf{r} \|} \dd S + \\
    &-\int_{\Gamma} \phi(\Tilde{\mathbf{r}}) \dfrac{e^{-k_D\| \Tilde{\mathbf{r}}-\mathbf{r} \|}(1+k_D\| \Tilde{\mathbf{r}}-\mathbf{r} \|)(\Tilde{\mathbf{r}}-\mathbf{r}) \cdot \mathbf{n}(\Tilde{\mathbf{r}})}{4\pi\| \Tilde{\mathbf{r}}-\mathbf{r} \|^3}  \dd S \; ,
\end{split}
% \label{eq:id_green_out2}
\end{equation*}
which actually yields Eq.~\eqref{eq:out_pot}.

\section{Details on analytical calculations done in Sect.~\ref{analytical_benchmark_section_label}}
\label{analytical_benchmark_appendix_label}
Unknown coefficients $L_{n l,i}$ and $G_{n l,i}$ of \eqref{Lin_pbe_phi_in_out} are to be determined from boundary conditions on the boundaries of spheres -- namely, 
\begin{equation}
\label{Lin_pbe_phi_bc}
\begin{aligned}
\left.\phi_{\text{in},i}\right|_{\varrho_i \to R_i^-} & = \left.\phi_{\text{out}}\right|_{\varrho_i \to R_i^+} , \\ 
\varepsilon_{r,m}\left.(\Hat{\mathbf n}_i\cdot\nabla\phi_{\text{in},i})\right|_{\varrho_i\to R_i^-} & = \varepsilon_{r,s}\left.(\Hat{\mathbf n}_i\cdot\nabla\phi_{\text{out}})\right|_{\varrho_i\to R_i^+}, 
\end{aligned}
\end{equation}
where $\Hat{\mathbf n}_i$ is the outer unit normal on the boundary of~$\Omega_{m,i}$. When imposing boundary conditions \eqref{Lin_pbe_phi_bc} on the $i$-th spherical surface, to have all quantities represented through the same basis functions set (namely, spherical harmonics $\{Y_n^l(\Hat{\boldsymbol\varrho}_i)\}_{0\le\left|l\right|\le n}$), we expand the offside (i.e.~those with respect to center $\mathbf r_j$, $j\ne i$) screened harmonics $\{k_n(\Tilde\varrho_j) Y_n^l(\Hat{\boldsymbol\varrho}_j)\}_{0\le\left|l\right|\le n}$ in $\phi_{\text{out}}$ using relations (see \cite{Yu3,Yu2021}) $k_L(\Tilde\varrho_j) Y_L^M(\Hat{\boldsymbol\varrho}_j) = \sum\nolimits_{l_1,m_1}\mathcal H_{l_1 m_1}^{L M}(\mathbf{a}_{i j}) i_{l_1}(\Tilde\varrho_i) Y_{l_1}^{m_1}(\Hat{\boldsymbol\varrho}_i)$ with coefficients 
$\mathcal H_{l_1 m_1}^{L M}(\mathbf{a}_{i j}) = \sum_{l_2,m_2} (-1)^{l_1+l_2} H_{l_1 m_1 l_2 m_2}^{L M} k_{l_2}(\Tilde a_{i j}) Y_{l_2}^{m_2}(\Hat{\mathbf a}_{i j})$ and $H_{l_1 m_1 l_2 m_2}^{L M} = C_{l_1 0 l_2 0}^{L 0} C_{l_1 m_1 l_2 m_2}^{L M} \sqrt{\frac{4\pi (2 l_1+1) (2 l_2+1)}{2 L+1}}$, where $\mathbf a_{i j}=\mathbf r_j-\mathbf r_i$ points from $\mathbf r_i$ to $\mathbf r_j$, $\varrho_i<a_{i j} = \|\mathbf a_{i j}\|$, $\Tilde a_{i j} = k_D a_{i j}$, $C_{l_1 m_1 l_2 m_2}^{L M} = \left<l_1 l_2; m_1 m_2 \mid L M\right>$ are Clebsch-Gordan coefficients, $i_n(\cdot)$ are modified spherical Bessel functions of the 1st kind. Doing so, we, in particular, ensure the correct mathematical treatment of mutual polarization effects \cite{our_jcp}. These operations readily convert boundary conditions \eqref{Lin_pbe_phi_bc} into a linear algebraic system governing the unknown coefficients of \eqref{Lin_pbe_phi_in_out}; by numerically solving it, unknown potentials \eqref{Lin_pbe_phi_in_out} are thereby completely determined. Note that in practical numerical calculations, in a manner completely similar to the particular case of 2-sphere systems extensively described in \cite{our_jcp}, one needs to limit index $n$ in \eqref{Lin_pbe_phi_in_out} (and this then naturally impacts the size of the linear system formed) by some user-defined threshold $n_\text{max}$, i.e.~one has $0\le n\le n_\text{max}$ everywhere in the calculations; now gradually increasing $n_\text{max}$ one ensures that potentials and/or other monitored quantities (such as energy) stop changing (their changes become negligible and not affecting the data within the precision reported in the work).

In order to establish equality \eqref{phi_pol_ii} we observe that the total polarization charge density at the $i$-th surface ($\varrho_i=R_i$) is~\cite{our_jcp}
\begin{align*}
& \sigma_{\text{tot},i}(\Hat{\boldsymbol\varrho}_i) = \varepsilon_0 \!\left(\frac{1}{\varepsilon_{r,m}} - \frac{1}{\varepsilon_{r,s}}\right)\!\varepsilon_{r,m} k_D \frac{\partial}{\partial\Tilde\varrho_i}\phi_{\text{in},i}\Bigr|_{\varrho_i=R_i} \\ 
& \  = \varepsilon_0 \!\left(\frac{1}{\varepsilon_{r,m}} - \frac{1}{\varepsilon_{r,s}}\right)\!\varepsilon_{r,m} k_D \biggl(\frac{- q_i k_D}{4\pi\varepsilon_0\varepsilon_{r,m} \Tilde\varrho_i^2} + \sum_{n,l} L_{n l,i} n \Tilde\varrho_i^{n-1} Y_n^l(\Hat{\boldsymbol\varrho}_i)\!\biggr)\biggr|_{\varrho_i=R_i}, 
\end{align*}
so that the corresponding resulting reaction potential $\phi_{\text{pol},i,i}$ created by this density at point $\mathbf r_i$ (the $i$-th sphere's center) is thus expressed by integral over the surface $\partial\Omega_{m,i}$ of sphere $\Omega_{m,i}$ (note the orthogonality property of spherical harmonics when integrating them over sphere): 
\begin{equation*}
\phi_{\text{pol},i,i} = \frac{1}{4\pi\varepsilon_0}\oint\nolimits_{\partial\Omega_{m,i}} \frac{\sigma_{\text{tot},i}(\Hat{\boldsymbol\varrho}_i)}{R_i} \dd S = \frac{q_i}{4\pi\varepsilon_0 R_i}\!\left( \frac{1}{\varepsilon_{r,s}} - \frac{1}{\varepsilon_{r,m}}\right) \!,
\end{equation*} 
that is equality~\eqref{phi_pol_ii}.

Next, in order to establish equality \eqref{phi_pol_ij}, i.e.~to evaluate the resulting potential $\phi_{\text{pol},i,j} = \frac{1}{4\pi\varepsilon_0}\oint_{\partial\Omega_{m,i}} \frac{\sigma_{\text{tot},i}(\Hat{\boldsymbol\varrho}_i) \dd S}{\|R_i \Hat{\boldsymbol\varrho}_i - \mathbf a_{i j}\|}$ created by density $\sigma_{\text{tot},i}$ at point $\mathbf r_j$ (note that $R_i \Hat{\boldsymbol\varrho}_i - \mathbf a_{i j}$ is the vector, $\boldsymbol\varrho_j$, pointing from $\mathbf r_j$ to the integration point on $\partial\Omega_{m,i}$ corresponding to $R_i \Hat{\boldsymbol\varrho}_i$) let us use the addition theorem for spherical harmonics (see \cite[\S~3.6]{Jack}) to express $$\frac{1}{\|R_i \Hat{\boldsymbol\varrho}_i - \mathbf a_{i j}\|} = \frac{4\pi}{a_{i j}}\sum_{n,l}\frac{1}{2 n+1} \biggl(\frac{R_i}{a_{i j}}\biggr)^{\!n} Y_n^l(\Hat{\mathbf a}_{i j}) Y_n^l(\Hat{\boldsymbol\varrho}_i)^\star,$$ from which taking into account the orthogonality property of spherical harmonics one then readily obtains relation~\eqref{phi_pol_ij}.